\documentclass[twocolumn,superscriptaddress,floatfix,preprintnumbers,letterpaper]{revtex4-1}
\usepackage[utf8]{inputenc}
\usepackage{array}
\usepackage{dcolumn}
\usepackage{bm}
\usepackage[colorlinks, pdfborder={0 0 0}, plainpages=false]{hyperref}
\usepackage{graphicx}
\usepackage{xspace}
\usepackage[usenames,dvipsnames]{color}
\usepackage{amssymb}
\usepackage[normalem]{ulem} 
\usepackage{lineno}
\usepackage{letltxmacro}
\usepackage{amsfonts}
\usepackage{amsmath}
\usepackage{amssymb}
\usepackage{bm}
\usepackage{dcolumn}
\usepackage{epsfig}
\usepackage{graphicx}
\usepackage{graphics}
\usepackage{latexsym}
\usepackage{rotating}
\usepackage{hyperref}
\usepackage[usenames]{color}
\usepackage{float}
\usepackage{xspace} 
\usepackage{mathrsfs}
\usepackage{subfigure}
\usepackage{multirow}
\usepackage{booktabs}
\usepackage{soul}

\usepackage{colortbl}
\usepackage{etoolbox}
\usepackage[table]{xcolor}

\def\prd{{\it Phys. Rev.} D~}

\def\apjl{{\it Astrophys. J.} Lett~}
\def\apj{{\it Astrophys. J.}}
\def\Msun{M_\odot}
\def\PRD{{\it Phys. Rev.} D~}

\def\lesssim{\mathrel{\hbox{\rlap{\hbox{\lower4pt\hbox{$\sim$}}}\hbox{$<$}}}}
\def\gtrsim{\mathrel{\hbox{\rlap{\hbox{\lower4pt\hbox{$\sim$}}}\hbox{$>$}}}}
\def\alt{\mathrel{\hbox{\rlap{\hbox{\lower4pt\hbox{$\sim$}}}\hbox{$<$}}}}
\def\agt{\mathrel{\hbox{\rlap{\hbox{\lower4pt\hbox{$\sim$}}}\hbox{$>$}}}}

\usepackage{mathtools}
\DeclarePairedDelimiter\abs{\lvert}{\rvert}

\newenvironment{cititemize2}
{\begin{list}{$\bullet$}
        {\setlength{\topsep}{0pt}
         \setlength{\itemsep}{0pt}
         \setlength{\parsep}{0.25\parsep}
         \settowidth{\labelwidth}{$\bullet$}
         \setlength{\leftmargin}{1em}
}
}
{\end{list}}

\def\gta{\ifmmode {\mathbin{\lower 3pt\hbox   
    {$\,\rlap{\raise 5pt\hbox{$\char'076$}}\mathchar"7218\,$}}}
    \else {${\mathbin{\lower 3pt\hbox
    {$\rlap{\raise 5pt\hbox{$\char'076$}}\mathchar"7218\,$}}}
    $}\fi}
\def\lta{\ifmmode {\,\mathbin{\lower 3pt\hbox   
    {$\,\rlap{\raise 5pt\hbox{$\char'074$}}\mathchar"7218\,$}}}
    \else {${\mathbin{\lower 3pt\hbox
    {$\rlap{\raise 5pt\hbox{$\char'074$}}\mathchar"7218\,$}}}
    $}\fi}

\newcommand{\msun}{{\rm M}_{\odot}}
\newcommand{\beq}{\begin{equation}}
\newcommand{\eeq}{\end{equation}}
\newcommand{\bea}{\begin{eqnarray}}
\newcommand{\eea}{\end{eqnarray}}

\definecolor{darkperiwinkle}{RGB}{102, 102, 128}

\newcommand{\NCSA}{\affiliation{NCSA, University of Illinois at Urbana-Champaign, Urbana, Illinois 61801, USA}}
\newcommand{\ANCSA}{\affiliation{Department of Astronomy, University of Illinois at Urbana-Champaign, Urbana, Illinois 61801, USA}}
\newcommand{\PNCSA}{\affiliation{Department of Physics, University of Illinois at Urbana-Champaign, Urbana, Illinois 61801, USA}}

\newcommand{\UNI}{\affiliation{The University of Illinois Laboratory High School, University of Illinois at Urbana-Champaign, Urbana, Illinois 61801, USA}}

\newcommand{\STAN}{\affiliation{Institute for Computational and Mathematical Engineering, Stanford University, Stanford, CA 94305, USA}}
\newcommand{\GoogleX}{\affiliation{Google X, Mountain View, California 94043, USA}}

\definecolor{light-gray}{gray}{0.9}

\begin{document}

\title{Fusing numerical relativity and deep learning \\ to detect higher-order multipole waveforms \\ from eccentric binary black hole mergers}

\author{Adam Rebei}\NCSA\UNI
\author{E. A. Huerta}\NCSA\ANCSA
\author{Sibo Wang}\NCSA
\author{Sarah Habib}\NCSA\PNCSA
\author{Roland Haas}\NCSA
\author{Daniel Johnson}\NCSA\STAN
\author{Daniel George}\NCSA\GoogleX

\date{\today}

\begin{abstract}
\noindent We determine the mass-ratio, eccentricity and binary inclination angles that maximize the contribution of the higher-order waveform multipoles \((\ell, \, \abs{m})= \{(2,\,2),\, (2,\,1),\, (3,\,3),\, (3,\,2), \, (3,\,1),\, (4,\,4),\, (4,\,3),\, (4,\,2),\,(4,\,1)\}\) for the gravitational wave detection of eccentric binary black hole mergers. We carry out this study using numerical relativity waveforms that describe non-spinning black hole binaries with mass-ratios \(1\leq q \leq 10\), and orbital eccentricities as high as \(e_0=0.18\) fifteen cycles before merger. For stellar-mass, asymmetric mass-ratio, binary black hole mergers, and assuming LIGO's Zero Detuned High Power configuration, we find that in regions of parameter space where black hole mergers modeled with \(\ell=\abs{m}=2\) waveforms have vanishing signal-to-noise ratios, the inclusion of \((\ell, \, \abs{m})\) modes enables the observation of these sources with signal-to-noise ratios that range between 30\% to 45\% the signal-to-noise ratio of optimally oriented binary black hole mergers modeled with \(\ell=\abs{m}=2\) numerical relativity waveforms. Having determined the parameter space where \((\ell, \, \abs{m})\) modes are important for gravitational wave detection, we construct waveform signals that describe these astrophysically motivate scenarios, and demonstrate that these topologically complex signals can be detected and characterized in real LIGO noise with deep learning algorithms.
\end{abstract}

\pacs{Valid PACS appear here}
\maketitle


\section{Introduction}
\label{intro}

\noindent The LIGO~\cite{DII:2016,LSC:2015} and Virgo~\cite{Virgo:2015} detectors have enabled the gravitational wave (GW) detection of binary black hole (BBH) mergers~\cite{DI:2016,secondBBH:2016,thirddetection,fourth:2017,GW170608,o1o2catalog} and neutron star (NS) collisions~\cite{bnsdet:2017}. These discoveries have firmly established GW astrophysics, and initiated the era of multi-messenger astrophysics (MMA)~\cite{bnsdet:2017,mma:2017arXiv,2017Sci...358.1556C}. 

In addition to providing a census of the mass and spin distributions of BHs and NSs, GWs can be used to infer information about the astrophysical environments where these objects form and coalesce. In this article, we focus on compact binary populations that may exist in dense stellar environments. The study of these potential GW sources has increased in recent years due to electromagnetic observations that are consistent with the existence of stellar mass BHs in the Galactic cluster M22~\cite{Strader:2012} and near the Galactic center~\cite{galcen:2018}. These observations have accelerated the development of algorithms to simulate the formation and retention of stellar mass BHs, and the formation of BBHs in dense stellar environments~\cite{antonini,Anton:2014,Antonini:2014,Anto:2015arXiv}. These improved algorithms have demonstrated that previous analyses significantly underestimated the detection rate of eccentric compact binary mergers, and have provided new insights into the expected eccentricity distribution of stellar-mass BH mergers in the frequency band of LIGO-type detectors~\cite{sam:2017ApJ...840L..14S,Samsing:2014,sam:171107452S,ssm:2017,ssm:2018,carl:171204937R}. 

Furthermore, we know from post-Newtonian calculations that the topology of GWs emitted by eccentric BBH mergers are remarkably different in nature to their quasi-circular counterparts. These differences are apparent even at the leading order post-Newtonian expansion of the waveform strain, which exhibits a much more complex dependence on the binary inclination angle. In view of these considerations, we quantify the importance of including higher-order waveform multipoles for the detection of eccentric BBH mergers. This study makes use of a catalog of numerical relativity (NR) waveforms that describe non-spinning BHs with mass-ratios \(1\leq q \leq 10\), and orbital eccentricities \(e_0\leq0.18\) fifteen cycles before merger~\cite{ecc_catalog}. 

As a preview of key results we present herein, we have found that the inclusion of higher-order waveform multipoles is significantly more important for eccentric BBH mergers than for their quasi-circular counterparts. For instance, in~\cite{Prayush:2013a} it was reported that the inclusion of higher-order waveform multipoles for quasi-circular, non-spinning, asymmetric mass-ratio BBH mergers could increase the signal-to-noise ratio (SNR) by as much as \(8\%\) in Advanced LIGO~\cite{ZDHP:2010}. In this study we report two significant findings.  First, the net effect of including higher-order modes is to help homogenize detectability of eccentric BBH mergers across the various orientations and positions they could have in the sky. Second, regions of parameter space in which \(\ell=\abs{m}=2\) signals have negligible SNRs, \((\ell, \,\abs{m})\) waveforms can be detected with SNRs that correspond to up to 45\% of the SNR of an optimally oriented \(\ell=\abs{m}=2\) eccentric BBH counterpart. We have also explored the dependance of these results on the initial eccentricity of the system, \(e_0\), for a wide range of mass-ratios. As in the case of quasi-circular, non-spinning, comparable mass-ratio BBH systems~\cite{Prayush:2013a}, we have found that the inclusion of higher-order waveform modes has a marginal impact in the detectability of these sources. However, as we show in this study, \((\ell, \,\abs{m})\) modes become significant for GW detection for BBH mergers with asymmetric mass-ratios. We also present results for the increase in the range of detectability of eccentric BBH mergers when the GW emission from these systems includes \((\ell, \,\abs{m})\) modes. 

Having constrained the parameter space that maximizes the contribution of \((\ell, \,\abs{m})\) for GW detection, we use these results as input information to explore the detectability of the corresponding higher-order waveform multipole signals in simulated and real LIGO noise. Since the NR waveforms we use for this study describe BBH mergers whose mass-ratios and eccentricities have not been previously studied in the literature~\cite{ecc_catalog}, these results provide new insights into the detectability of these types of signals in simulated and real LIGO noise.

It is worth pointing out that most of the existing tools for GW detection are tailored for the identification of quasi-circular binaries~\cite{gstlal:2012ApJ,PhysRevD.95.042001,2016CQGra..33u5004U}. To the best of our knowledge, no matched-filtering algorithm has been presented in the literature for the detection of eccentric BBH mergers~\cite{Tiwari:2016}. On the other hand, while template agnostic searches have been utilized to search for eccentric BBH mergers~\cite{Tiwari:2016,Sergey:2016}, this class of algorithms is optimal for the detection of burst-like GW signals, and may miss \({\cal{O}}(\textrm{second-long})\) signals with low SNRs~\cite{secondBBH:2016}.

A different class of signal-processing algorithm, based on deep neural networks, was pioneered by authors in this manuscript for the detection of GWs in simulated and real LIGO noise~\cite{geodf:2017a,geodf:2017b}. These algorithms have been used to identify, reconstruct  and denoise eccentric NR waveforms that contain the dominant \(\ell=\abs{m}=2\) mode embedded both in simulated and real LIGO noise~\cite{geodf:2017a,geodf:2017b,geodf:2017c,hshen:2017,wei:2019W,dgNIPS}. Authors of this manuscript have also designed the first generation of deep learning algorithms at scale~\cite{shen_dlscale:2019} to estimate the parameters of the catalog of BBH mergers detected by the advanced LIGO and Virgo detectors~\cite{o1o2catalog}. In view of these developments, and realizing that the vast majority of eccentric waveform models have only used the \(\ell=\abs{m}=2\) mode~\cite{cao:2017,Hinderer:2017,Huerta:2017a,huerta:2018PhRvD,Levin:2011C,Yunes:2009,Huerta:2014,Huerta:2013a,Hinder:2010,lou:2016arXiv,lou:2014PhRvD,Osburn:2016,lou:2017CQG,moore:2018fde}, in this article we use NR waveforms to quantify the importance of including higher-order waveform multipoles for the GW detection of eccentric BBH mergers, and demonstrate that deep learning can identify these complex signals both in simulated and raw advanced LIGO noise. 

This article is organized as follows: Section~\ref{nr_cat} describes the NR waveforms used for this study, and the method used to construct higher-order waveform multipole signals. In Section~\ref{ho_modes} we constrain the parameter space that maximizes the contribution of \((\ell, \,\abs{m})\) for GW detection. We use these results as input data to compute the SNR distribution of astrophysically motivated BBH populations in Section~\ref{sec:snrs}. In Section~\ref{range} we quantify the importance of including \((\ell, \,\abs{m})\) modes on the range of detectability for eccentric BBH mergers. Section~\ref{ecc_det} introduces deep learning algorithms to detect eccentric higher-order waveform multipole signals in simulated and real LIGO noise. We summarize our findings and future directions of work in Section~\ref{end}.


\section{Numerical relativity catalog} 
\label{nr_cat}

We use a catalog of NR waveforms~\cite{ecc_catalog}, produced with the  \texttt{Einstein Toolkit}~\cite{naka:1987,shiba:1995,baum:1998,baker:2006,camp:2006,Lama:2011,wardell_barry_2016_155394,ETL:2012CQGra,Ansorg:2004ds,Diener:2005tn,Schnetter:2003rb,Thornburg:2003sf} that describe BBHs with mass-ratios \(1\leq q \leq 10\), and eccentricities up to \(e_0\leq0.18\) fifteen cycles before merger. We extracted the modes  \((\ell, \, \abs{m})= \{(2,\,2),\, (2,\,1),\, (3,\,3),\, (3,\,2), \, (3,\,1),\, (4,\,4),\, (4,\,3),\, (4,\,2)\), \((4,\,1)\}\) using the open source software \texttt{POWER}~\cite{johnson:2017}. 

Given that the orbital eccentricity is a gauge-dependent parameter, and the tracks of the BHs in the NR simulations do not provide a measurement of the orbital eccentricity due to its gauge-dependent nature, it is necessary to use a gauge-independent method to obtain a measurement of the orbital eccentricity and mean anomaly of our waveform catalog. We have used the inspiral-merger-ringdown eccentric waveform model \texttt{ENIGMA}~\cite{huerta:2018PhRvD} to measure the initial eccentricity, \(e_0\), and mean anomaly, \(\ell_0\), at a time \(t_0\) when the NR waveforms are free from junk radiation~\cite{Habib_Huerta_2019a}.  

The \texttt{ENIGMA} waveform model uses the following parameterization 

\begin{equation} 
e = e(x)\,, \quad \ell =  \ell(x)\,, \quad x=\left(M \omega \right)^{2/3}\,,
\end{equation}

\noindent where \(M\) stands for the total mass of the BBH, and \(\omega\) represents the mean orbital frequency. We choose this approach because post-Newtonian approximations parameterized in terms of the gauge-invariant parameter, \(x\), reproduce with better accuracy the dynamics of eccentric BBH mergers, as compared to NR simulations~\cite{ian:2017,Hinder:2010,Huerta:2017a,huerta:2018PhRvD}. Therefore, upon integrating the equations of motion of the BBH system, given by equations~(1)-(6) in~\cite{Huerta:2017a}, we can solve for \(e(t=t_0)=e_0\), \(\ell(t=t_0)=\ell_0\) and \(x(t=t_0)=x_0\). Using this scheme, we construct the \texttt{ENIGMA} waveforms with the optimal value for the triplet \((e_0,\, \ell_0,\,x_0)\) that maximize the overlap with NR waveforms that represent the same BBH systems. These parameters fully describe our NR waveforms, and we present them in Table~\ref{results}. Having extracted the \((\ell\,,\abs{m})\) modes listed above, we construct the  waveform strain \(h(t,\, \theta,\,\phi)\) using the extracted \(h^{\ell m}\) waveforms and the spin-weight--2 spherical harmonics, \({}_{-2}Y_{\ell m} \left(\theta, \phi\right)\)~\cite{Blanchet:2006}.

\begin{equation}
h(t,\, \theta,\,\phi) = h_{+} - \textrm{i} h_{\times} = \sum_{\ell\geq2}\,\,\sum_{m \geq -\ell}^{m \leq \ell}h^{\ell m}{}_{-2}Y_{\ell m} \left(\theta, \phi\right)\,,
\label{strain}
\end{equation} 

\noindent where the reference frame \((\theta\,,\phi)\) is anchored at the center of mass of the BBH, and determines the location of the GW detector. In this reference frame \(\theta=0\) coincides with the total angular momentum of the binary, and \(\phi\) indicates the azimuthal direction to the observer. On the other hand, the observed GW signal, \(H\), and the corresponding optimal matched-filter \(\textrm{SNR}\) are given by

\begin{align}
\label{full_strain}
H &= F_{+}h_{+}\left(\alpha,\,\beta\right) + F_{\times}h_{\times}\left(\alpha,\,\beta\right)\,,\\
\label{snr}
\textrm{SNR}^2 &= 4 \Re\int_{f_0}^{f_{\mathrm{max}}} \frac{\tilde{H}\tilde{H}^{*}}{S_n(f)}\mathrm{d}f\,,
\end{align}
 
\noindent \(f_0=10\mathrm{Hz}\) corresponds to the low frequency end of LIGO's sensitivity band, and \(f_{\mathrm{max}}=8192 \mathrm{Hz}\) is used to capture the SNR contribution from the merger of all the BBHs considered in this study. \(\tilde{H}\) represents the Fourier transform of \(H\); \((F_{+}, \, F_{\times})\) are the antenna pattern response functions which depend on the sky position \((\alpha,\, \beta)\) of GW sources, and the polarization angle \(\psi\)~\cite{Finn:2001}. \(S_n(f)\) is the one-sided noise power spectral density (PSD) corresponding to LIGO's Zero Detuned High Power configuration (ZDHP)~\cite{ZDHP:2010}. 

 \begin{table}
\caption{\label{results} The table presents the parameters of the binary black holes considered in this study, namely: \(q\) represents the mass-ratio of the binary components, whereas \((e_0,\, \ell_0)\) are the values of eccentricity and mean anomaly of the binaries' orbit measured at the dimensionless orbital frequency \(x_0\).}
		\footnotesize
		\begin{center}
                        \setlength{\tabcolsep}{12pt} 
			\begin{tabular}{c c c c c c}
				\hline 
				Simulation&$q$ & $e_0$ & $\ell_0$ & $x_0$ \\ 
				\hline
				E0001	&	1	&	0.052	&	3.0	&	0.0770	\\
				J0040	&	1	&	0.160	&	3.0	&	0.0761	\\
 				J0045	&	2	&	0.056	&	3.0	&	0.0793	\\
				J0061	&	4	&	0.060	&	3.0	&	0.0855	\\
				L0016	&	5	&	0.140	&	2.9	&	0.0862	\\
				P0001	&	6	&	0.050	&	3.0	&	0.0867	\\
				P0016	&	6	&	0.160	&	2.8	&	0.0900	\\
				P0006	&	8	&	0.080	&	2.9	&	0.0931	\\
				P0008	&	8	&	0.140	&	2.9	&	0.0910	\\
				P0017	&	8	&	0.060	&	3.0	&	0.0927	\\
				P0020	&	8	&	0.180	&	2.9	&	0.0936	\\
				P0009	&	10	&	0.060	&	2.9	&	0.0971	\\
				P0024	&	10	&	0.180	&	3.0	&	0.0957	\\
				\hline 
			\end{tabular}
		\end{center}
	\label{sims}
	\end{table}
	\normalsize


\section{Higher-order waveform multipoles} 
\label{ho_modes}

\begin{figure}
\centerline{
\includegraphics[width=\linewidth]{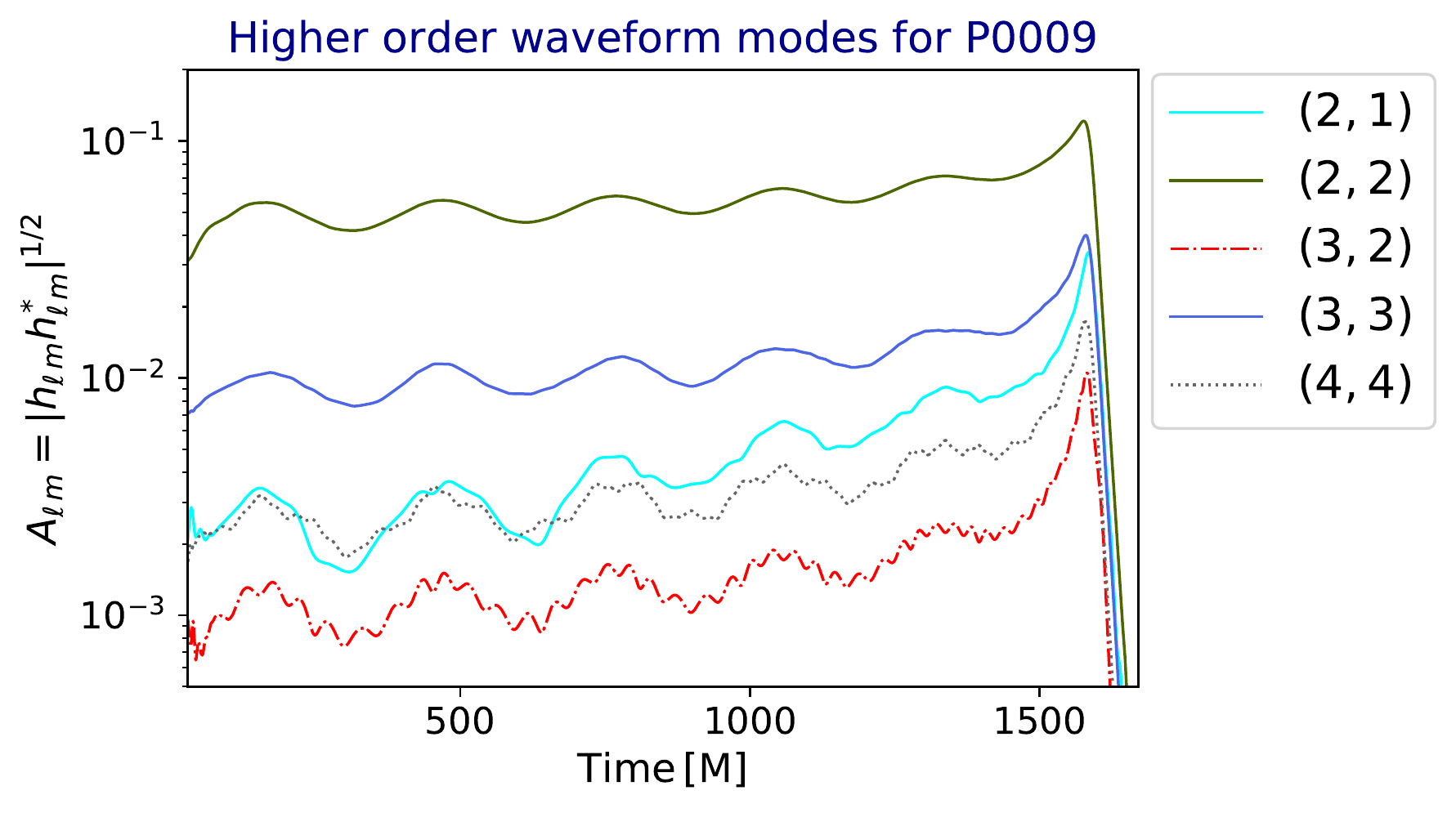}
}
\caption{Amplitude, \(A_{\ell m}= \left| h_{\ell m}h^{*}_{\ell m}\right|^{1/2}\), of the most significant \((\ell\,,\abs{m})\) modes for the NR simulation P0009.}
 \label{modes_P0009}
 \end{figure}

To motivate the first approach we have used to quantify the importance of including \((\ell\,,\abs{m})\) modes for GW detection, Figure~\ref{modes_P0009} presents the amplitude of the leading higher-order waveform modes for the NR simulation P0009 (see Table~\ref{sims}). 

\noindent At first glance, Figure~\ref{modes_P0009} suggests that higher-order modes will contribute significantly to the waveform strain \(h(t,\, \theta,\,\phi)\), see Eq.~\eqref{strain}, in the vicinity of merger. In view of this observation, our first study consists of constraining the \((\theta,\,\phi)\) parameter space that maximizes the contribution of \((\ell\,,\abs{m})\) modes near merger. Another complementary study we have conducted consists of determining the \((\theta,\,\phi)\) parameter space in which the higher-order modes contribute significantly throughout the entire waveform signal. We do so by using a metric that is similar in nature to the matched-filtering SNR calculation in Eq.~\ref{snr}. In the following subsections we describe each method, and provide results for both approaches.

\subsection{Contribution of higher-order modes near merger}
\label{metric_a_maps}

For our first study, we determine the \(\left(\theta,\,\phi\right)\) regions that maximize the contribution of \((\ell, \, \abs{m})\) in the vicinity of merger. In practice, we evaluate the waveform strain \(h\left(t,\,\theta,\,\phi\right)\), see Equation~\eqref{strain}, densely covering the \(\left(\theta,\,\phi\right)\) parameter space. Thereafter, we compute the time \((t^*, \hat{t})\) at which each \(h\left(t,\theta,\,\phi\right)\) attains its maximum value, and use it to compute the (scalar) waveform amplitude maximum \({\cal{A}}^{\left(\ell, \, \abs{m}\right)}(\theta,\,\phi)\)

\begin{eqnarray}
\label{max_all}
{\cal{A}}^{\left(\ell, \, \abs{m}\right)}(\theta,\,\phi) & = &\sqrt{h(t^*,\, \theta,\,\phi)\, \tilde{h}(t^*,\, \theta,\,\phi)}\,,\\
\label{max_two}
{\cal{A}}^{\left(\ell=\abs{m}=2\right)}(\theta,\,\phi) & = &\sqrt{h(\hat{t},\, \theta,\,\phi)\, \tilde{h}(\hat{t},\, \theta,\,\phi)}\,.
\end{eqnarray}

\begin{figure*}
\centerline{
\includegraphics[width=0.34\textwidth]{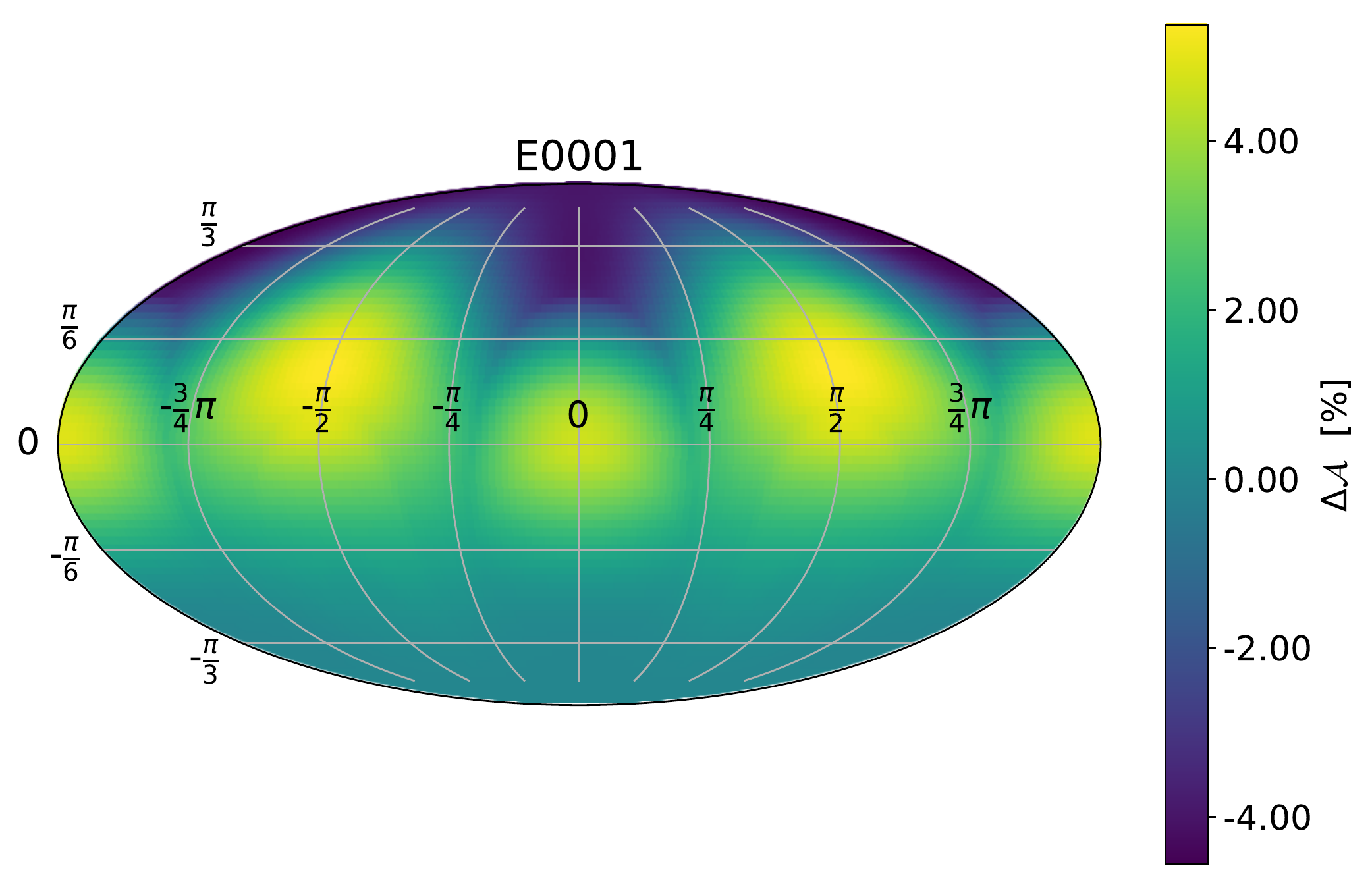}\hspace{0.1em}%
\includegraphics[width=0.34\textwidth]{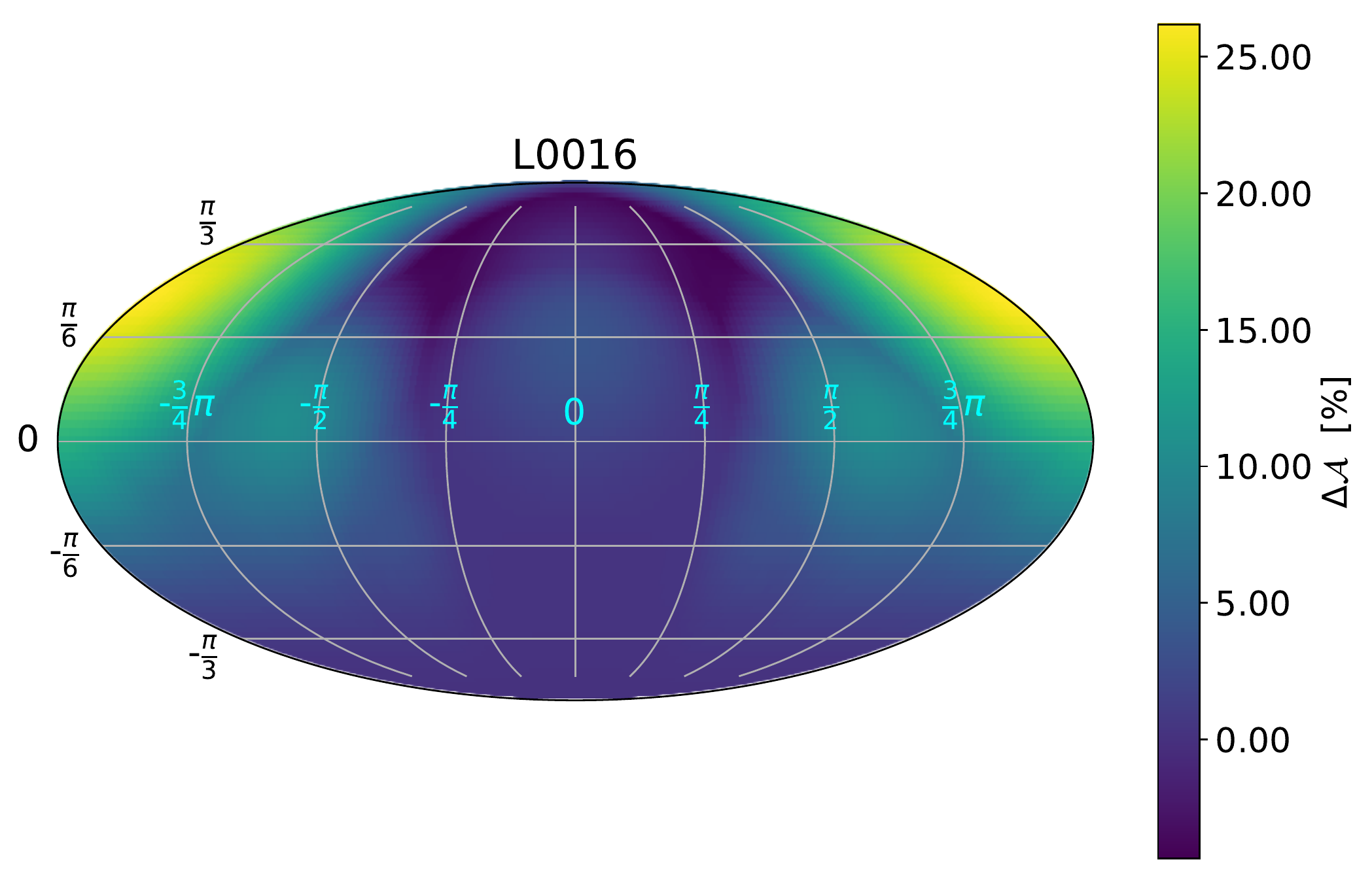}\hspace{0.1em}%
\includegraphics[width=0.34\textwidth]{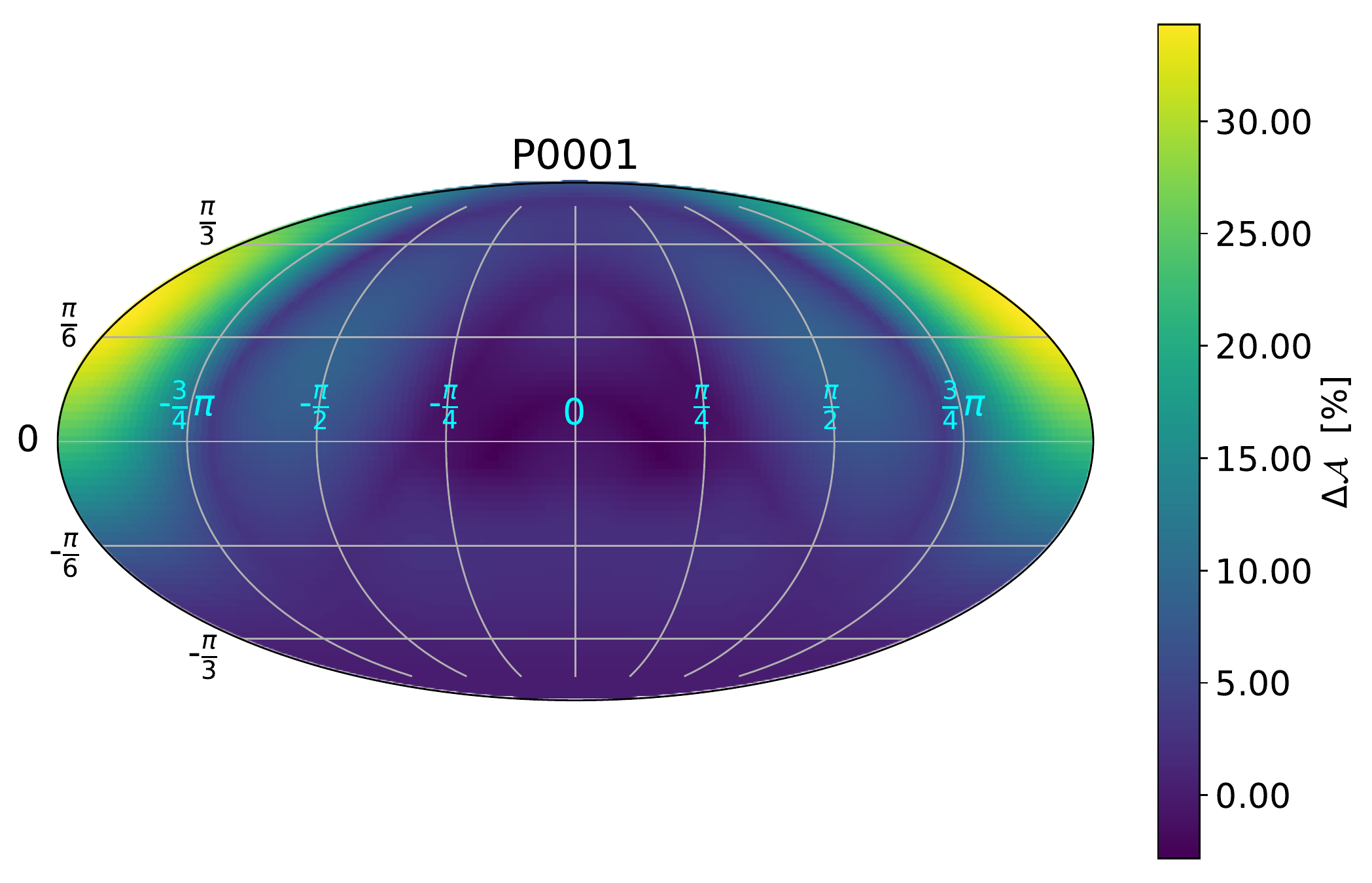}\hspace{0.1em}%
}
\centerline{
\includegraphics[width=0.34\textwidth]{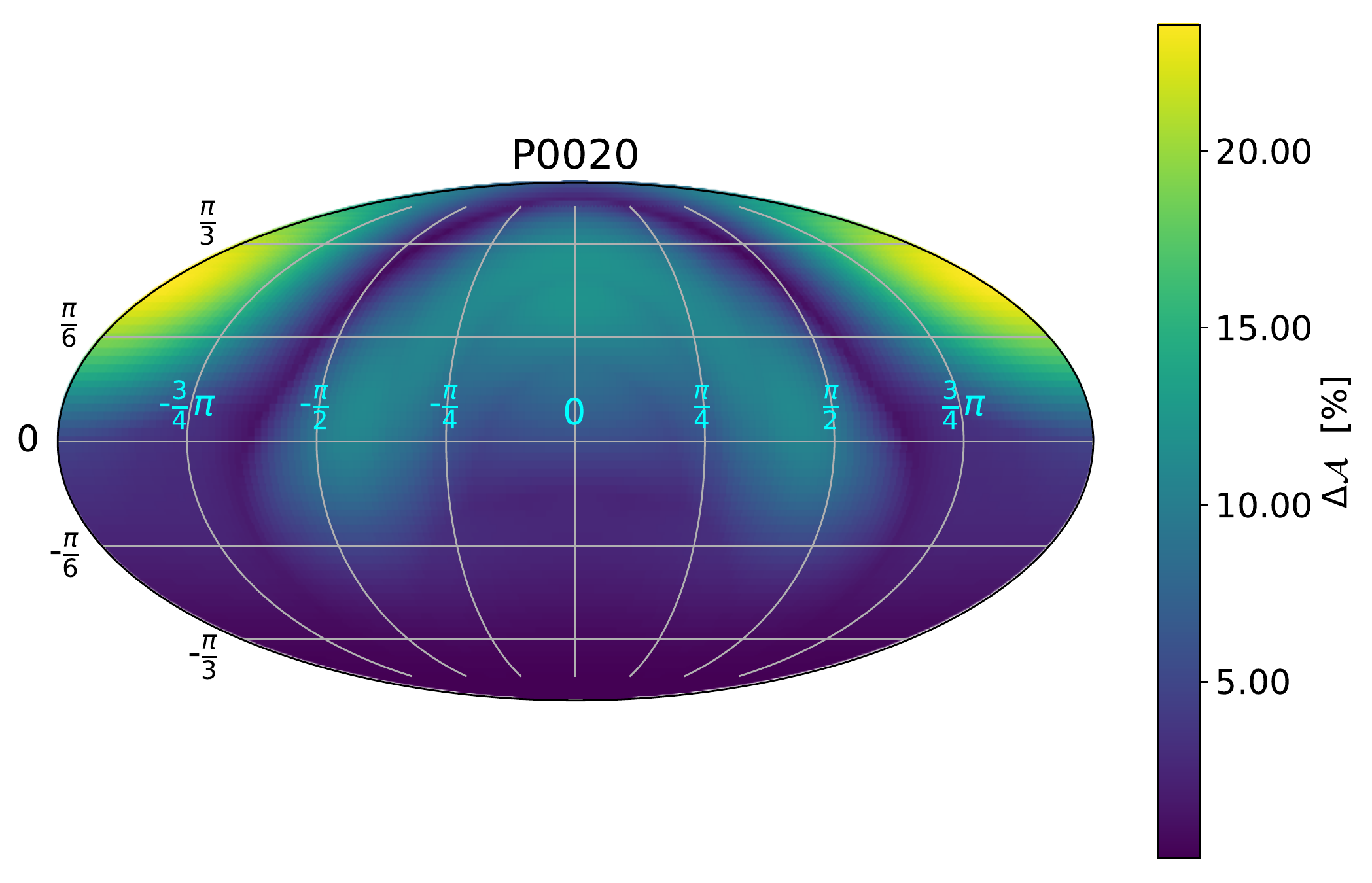}\hspace{0.9em}%
\includegraphics[width=0.34\textwidth]{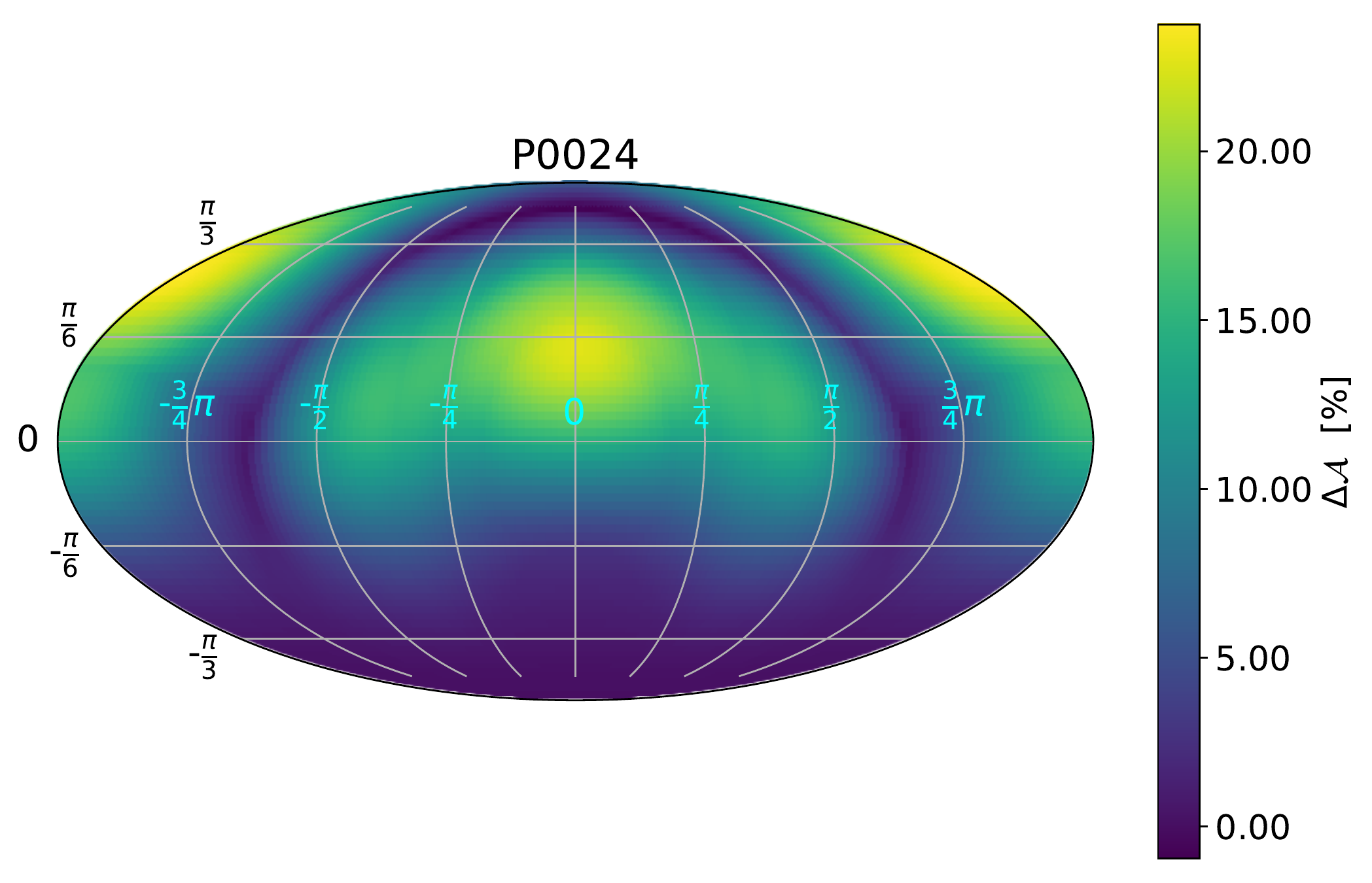}
}
\caption{\(\Delta {\cal{A}}(\theta,\,\phi)\), as defined in Eq.~\ref{a_difference}, constrains the regions of \( (\theta,\,\phi)\) parameter space that maximize the contribution of \((\ell, \, \abs{m})\) modes to the waveform amplitude near merger.  These \( (\theta,\,\phi)\) maps are constructed using the Mollweide projection: \((\vartheta, \varphi) \rightarrow (\pi/2-\theta, \phi-\pi)\).} 
 \label{maximization_A}
 \end{figure*}

\noindent Note that we have taken into account that waveforms that include \((\ell,\,\abs{m})\) modes or just the \(\ell=\abs{m}=2\) mode have different behavior in the vicinity of merger, in particular their maximum amplitude occurs at different times, \((t^*, \hat{t})\), respectively. Since we are just interested in computing the amplitude peak of both classes of waveforms, we compute the scalar quantity 

\begin{equation}
\Delta {\cal{A}}(\theta,\,\phi) = \frac{{\cal{A}}^{\left(\ell, \, \abs{m}\right)}(\theta,\,\phi) - {\cal{A}}^{\left(\ell=\abs{m}=2\right)}(\theta,\,\phi)}{{\cal{A}}^{\left(\ell=\abs{m}=2\right)}(\hat{\theta},\,\hat{\phi})}\,,
\label{a_difference}
\end{equation}

\noindent which measures, at each \((\theta,\,\phi)\), the waveform amplitude peak difference between NR waveforms that include all \((\ell, \, \abs{m})\) modes and those that only include the \(\ell=\abs{m}=2\) mode. We normalize this amplitude peak difference with respect to the maximum value of \({\cal{A}}^{\left(\ell=\abs{m}=2\right)}(\theta,\,\phi)\) across the \((\theta,\,\phi)\) parameter space, i.e., \({\cal{A}}^{\left(\ell=\abs{m}=2\right)}(\hat{\theta},\,\hat{\phi})\). The corresponding \((\theta,\,\phi)\) maps produced using the metric \(\Delta {\cal{A}}(\theta,\,\phi)\) are presented in Figure~\ref{maximization_A}. 

In the panels of Figure~\ref{maximization_A}, the bright yellow regions indicate the \((\theta,\,\phi)\) combinations that produce higher-order mode waveforms whose amplitude near merger is larger than waveforms that only include the \(\ell=\abs{m}=2\) mode. It is worth mentioning that the amplitude of \(\ell=\abs{m}=2\) waveforms is always maximized at the north pole of these Mollweide projections. In view of this observation, it is easy to understand that higher-order modes for equal-mass BBH binaries (top-left panel of Figure~\ref{maximization_A}) do not significantly contribute to the overall waveform amplitude since they are maximized near the equator of the Mollweide projections. However, as the mass-ratio and eccentricity increase, higher-order modes are maximized in \((\theta,\,\phi)\) regions that are closer to the maximum amplitude of the \(\ell=\abs{m}=2\) mode, as shown in the bottom-right panel of Figure~\ref{maximization_A}. At a glance, these results indicate that the \((\theta,\,\phi)\) regions where \((\ell, \, \abs{m})\) modes are maximized tend to shift towards the north pole as the mass-ratio and eccentricity increase.

\subsection{Contribution of higher-order modes throughout the waveform evolution}

The second study is inspired by a frequency to time-domain manipulation of the optimal SNR of a waveform signal \(\mathbf{h}\) that is filtered by a template \(\mathbf{T-h}\). The optimal SNR of a waveform signal \(\mathbf{h}\) is given as~\cite{cutler}

\begin{equation}
\label{snr_white}
\textrm{SNR}[\mathbf{h}]^2 = \left(\mathbf{h} | \mathbf{h} \right)=4 \Re\int_{0}^{\infty} \frac{\tilde{h}(f)\tilde{h}^{*}(f)}{S_n(f)}\mathrm{d}f\,.
\end{equation}

\begin{figure*}
\centerline{
\includegraphics[width=0.34\textwidth]{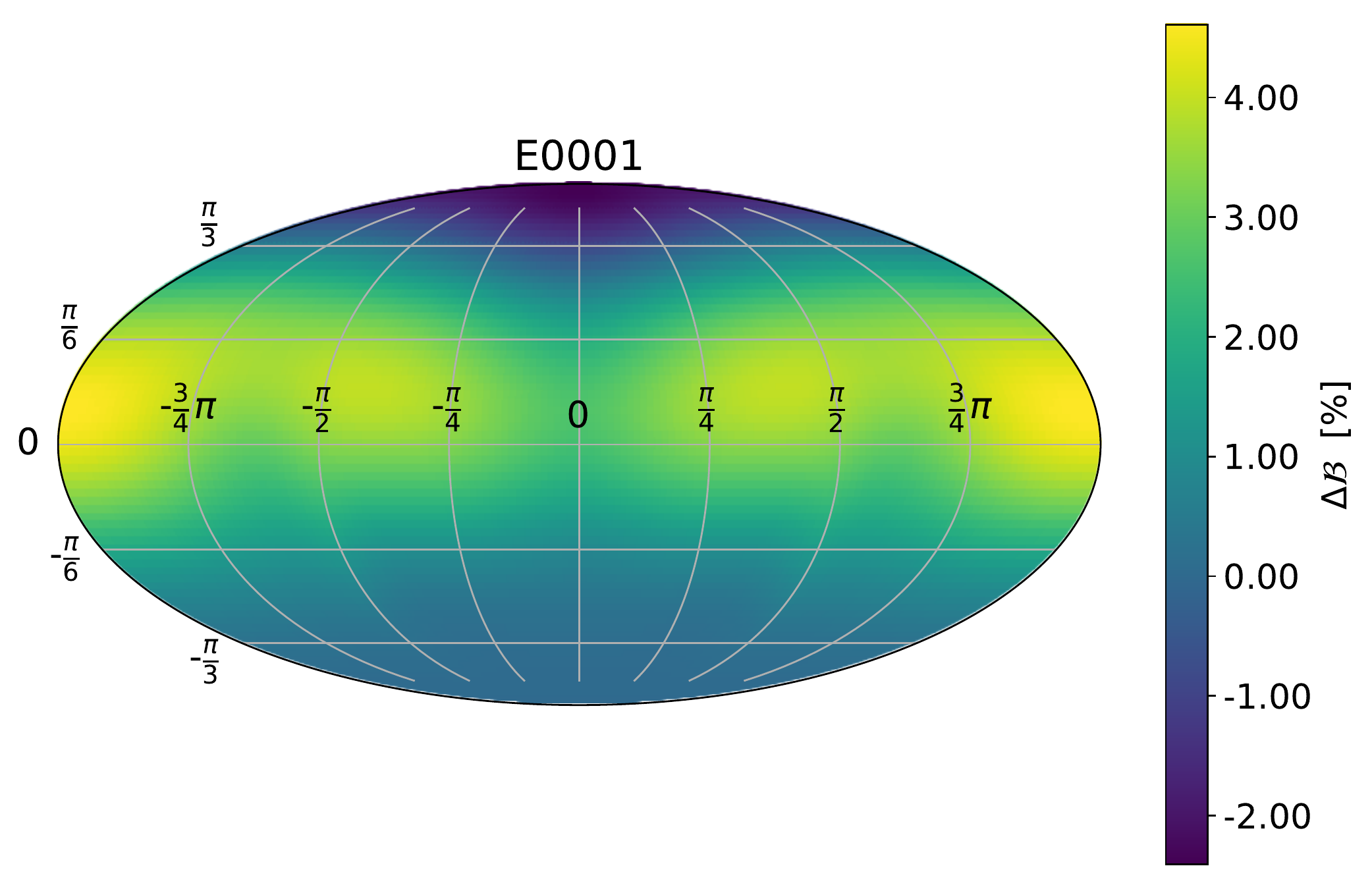}\hspace{0.1em}%
\includegraphics[width=0.34\textwidth]{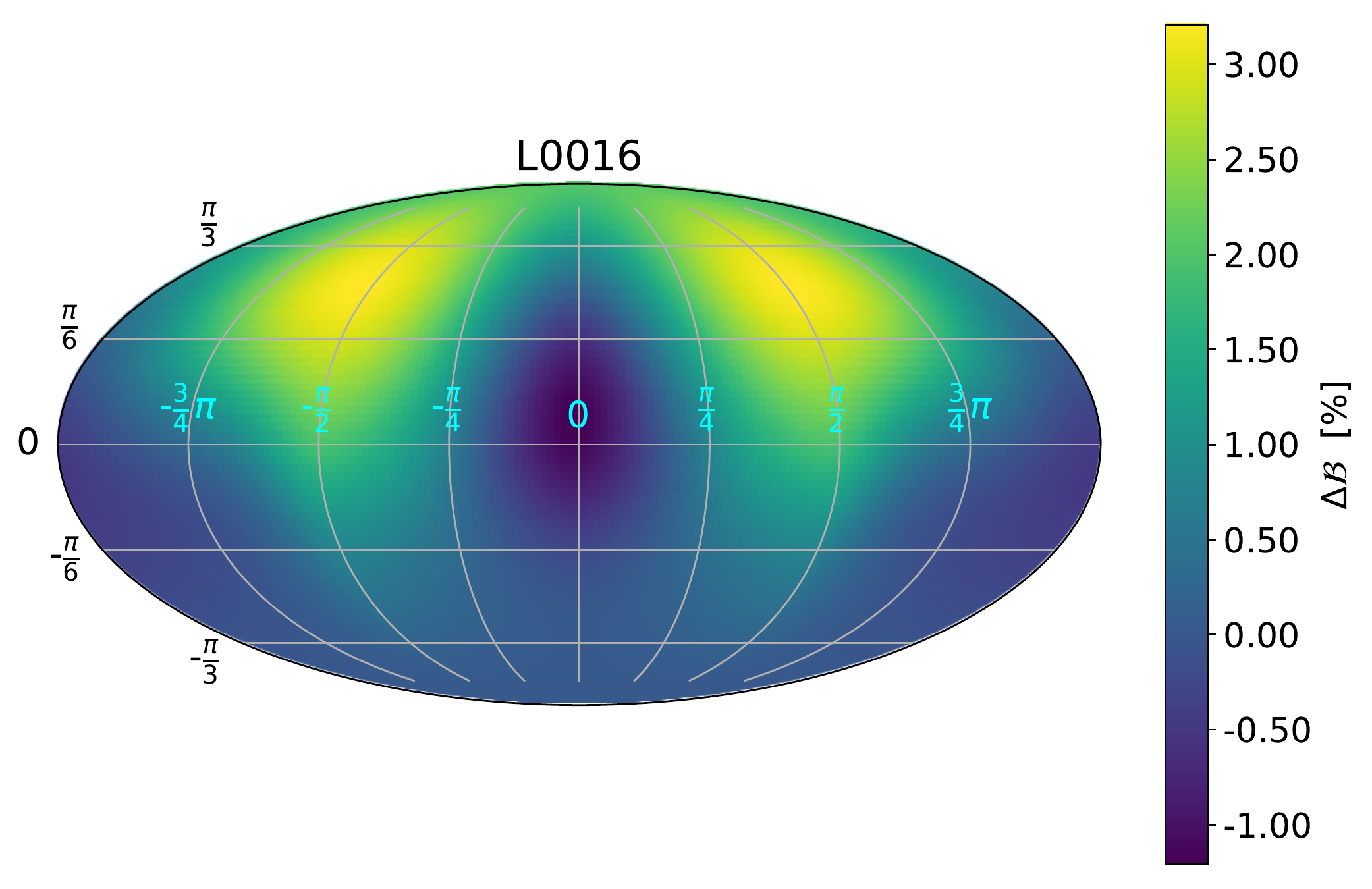}\hspace{0.1em}
\includegraphics[width=0.34\textwidth]{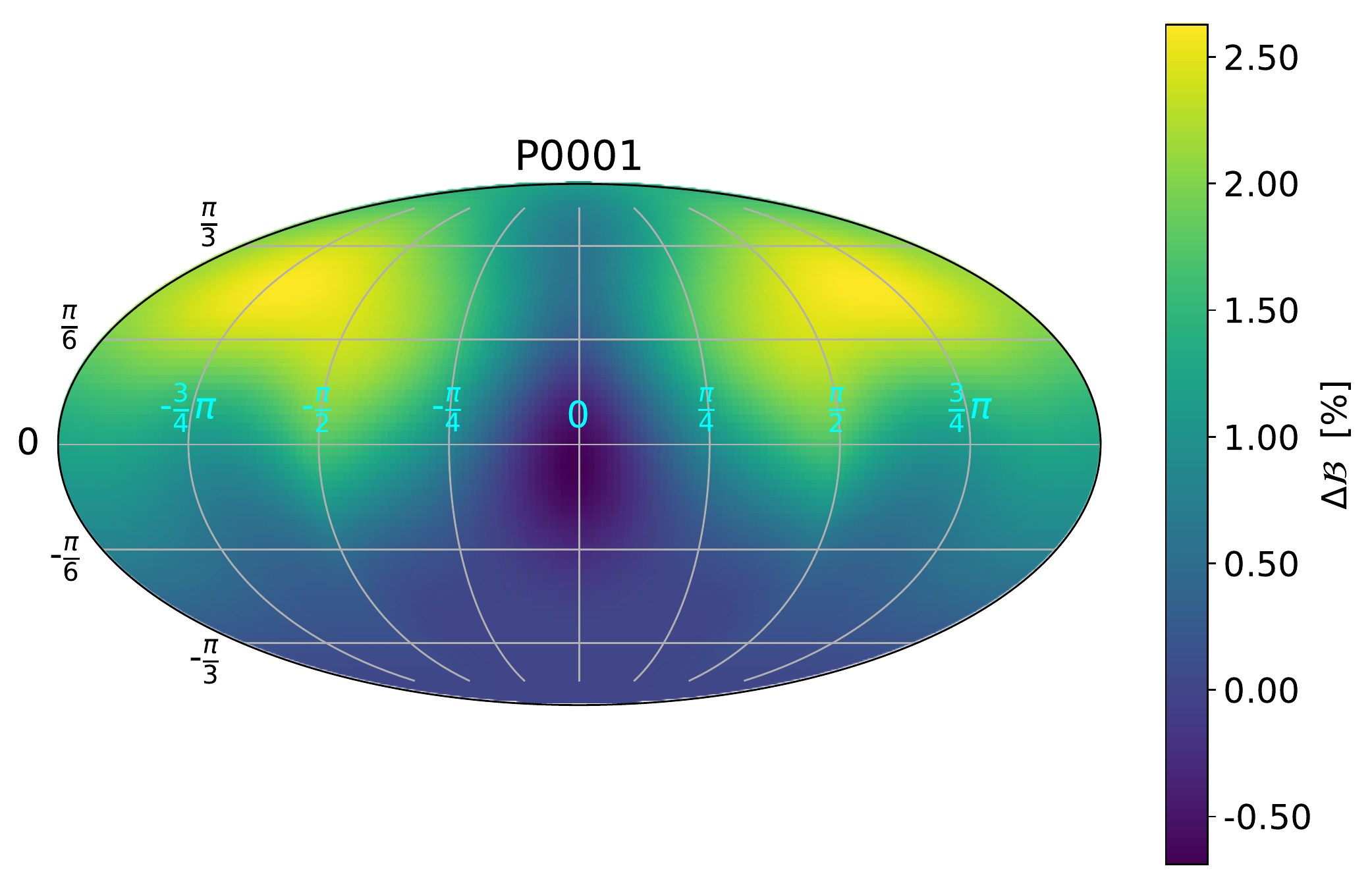}\hspace{0.1em}%
}
\centerline{
\includegraphics[width=0.34\textwidth]{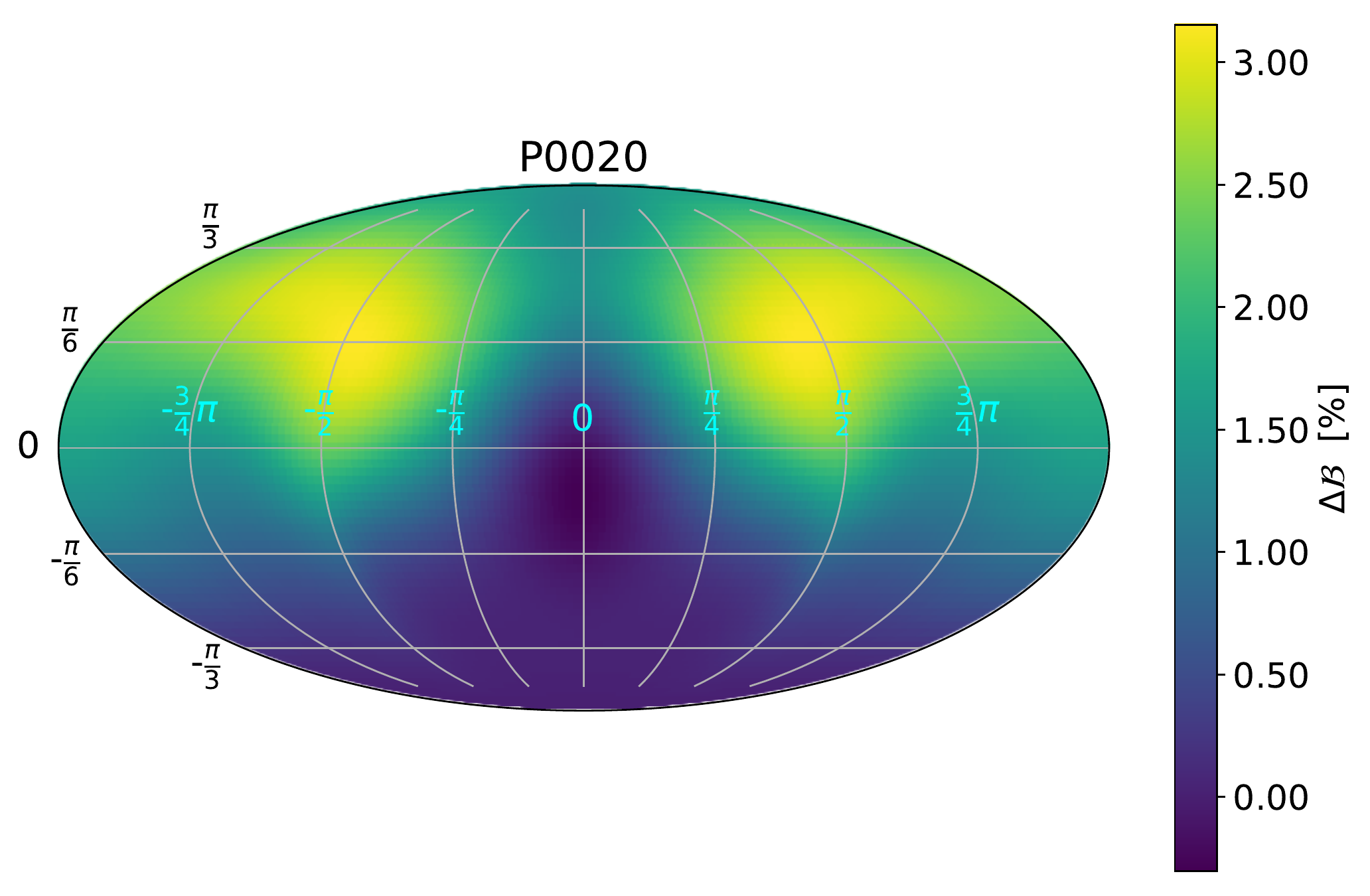}\hspace{0.9em}%
\includegraphics[width=0.34\textwidth]{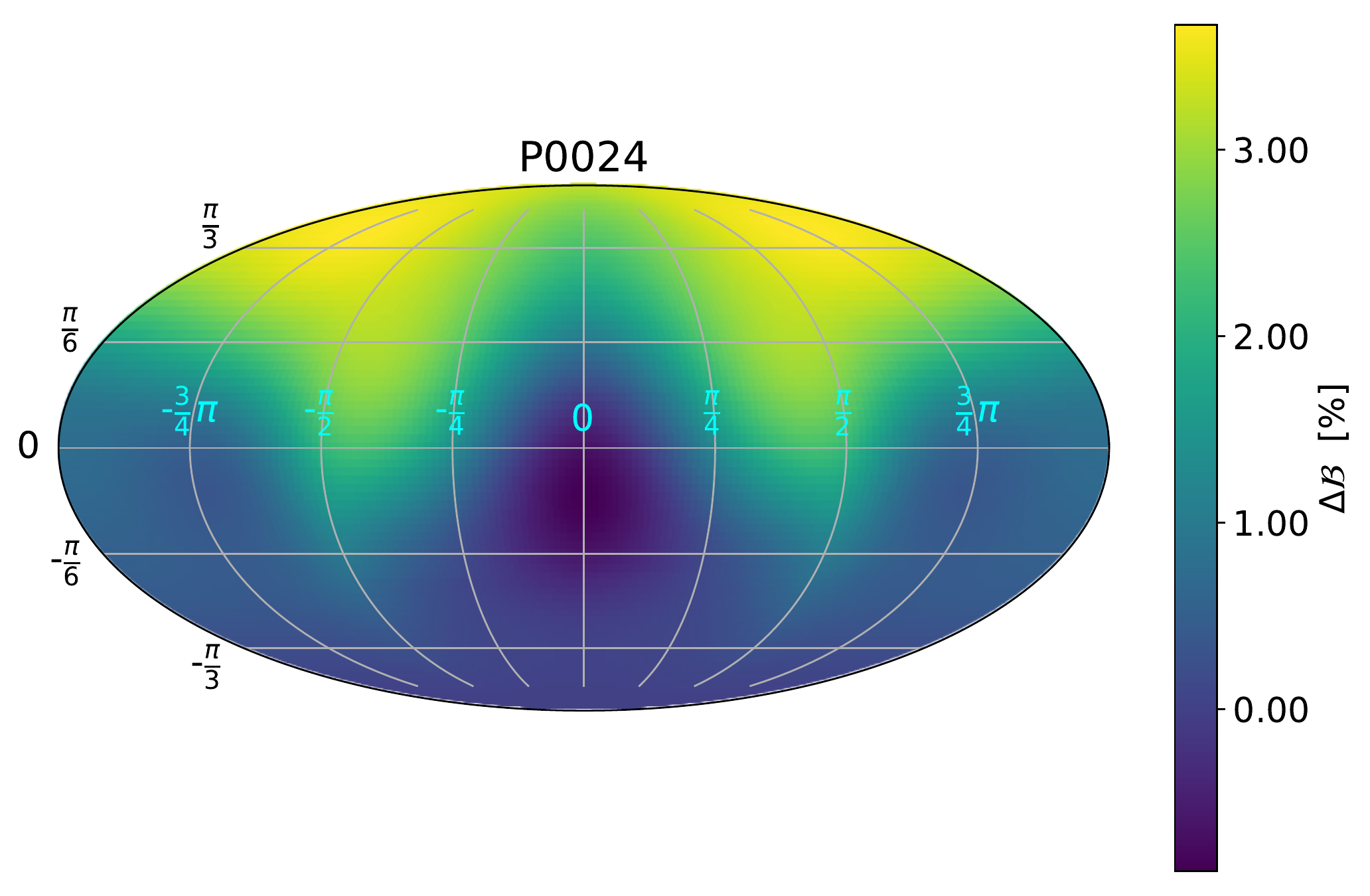}
}
\caption{\(\Delta {\cal{B}}\), as defined in Eq.~\eqref{area}, identifies the \( (\theta,\,\phi)\) regions where the integrated amplitude of NR waveforms that include \((\ell, \, \abs{m})\) modes is greater than than their \(\ell=\abs{m}=2\) counterparts. These \( (\theta,\,\phi)\) maps are constructed using the Mollweide projection: \((\vartheta, \varphi) \rightarrow (\pi/2-\theta, \phi-\pi)\).}
 \label{maximization}
 \end{figure*}

\noindent Additionally, assuming white noise, \(S_n(f) \approx \textrm{constant} \equiv S_0\). Using Parseval's theorem, one can recast Eq.~\eqref{snr_white} in time-domain as follows

\begin{equation}
\label{snr_white_time}
\textrm{SNR}[\mathbf{h}]^2 = \frac{2}{S_0} \int_{t=t_0}^{T} h(t)^2 \mathrm{d}t\,,
\end{equation}

\noindent where \(t_0\) represents the time from which the NR waveform is free from junk radiation, and \(T\) is the final time sample of the NR waveform. Using this time-domain expression of the SNR as motivation, we have densely sampled the \((\theta,\,\phi)\) parameter space, and computed the integrated amplitude of NR waveforms that include either \((\ell, \, \abs{m})\) modes, or just the \(\ell=\abs{m}=2\) mode, using the metric

\begin{eqnarray}
\label{int}
{\cal{B}}^{\left(\ell, \, \abs{m}\right)}(\theta,\,\phi) &=& \int_{t=t_0}^{T}  \sqrt{h(t,\, \theta,\,\phi)\, \tilde{h}(t,\, \theta,\,\phi)}\,\mathrm{d}t\,,\\
\label{area}
\Delta {\cal{B}}(\theta,\,\phi) &=& \frac{{\cal{B}}^{\left(\ell, \, \abs{m}\right)}\left(\theta,\,\phi\right)    - {\cal{B}}^{\left(\ell=\abs{m}=2\right)}\left(\theta,\,\phi\right)}{{\cal{B}}^{\left(\ell=\abs{m}=2\right)}(\hat{\theta},\,\hat{\phi})}\,,\nonumber\\
\end{eqnarray}

\noindent where \({\cal{B}}^{\left(\ell=\abs{m}=2\right)}(\hat{\theta},\,\hat{\phi})\) represents the maximum of 
\({\cal{B}}^{\left(\ell=\abs{m}=2\right)}\left(\theta,\,\phi\right)\) across the  \((\theta,\,\phi)\) parameter space. 
The maxima of \(\Delta {\cal{B}}^{\left(\ell, \, \abs{m}\right)}(\theta,\,\phi)\) 
identify the \((\theta,\,\phi)\) regions where the integrated amplitude of NR 
waveforms that include \((\ell, \, \abs{m})\) modes is greater than 
their \(\ell=\abs{m}=2\) counterparts. The \((\theta,\,\phi)\) maps for this metric are shown 
in Figure~\ref{maximization}. Note that since the vacuum spacetime NR 
waveforms we use are mass invariant, these results will remain valid 
for ground-based or space-based GW 
detectors, so long as the BBH merger takes place in the sensitive 
frequency band of the GW detector under consideration.

A direct comparison between the \(\Delta {\cal{A}}^{\left(\ell, \, \abs{m}\right)}(\theta,\,\phi) \) and 
\(\Delta {\cal{B}}^{\left(\ell, \, \abs{m}\right)}(\theta,\,\phi) \) metrics shows that they constrain 
different \((\theta,\,\phi)\) regions of parameter 
space. We also notice that the regions constrained by metric \(\Delta {\cal{B}}\) are much closer 
to the north pole, compared to metric \(\Delta {\cal{A}}\), as the mass-ratio and eccentricity of the 
BBHs increase. In different words, by using metric \(\Delta {\cal{B}}\), we have identified 
\((\theta,\,\phi)\) regions where higher-order modes significantly contribute to the integrated amplitude 
of waveform signals, and which are as significant as optimally oriented \(\ell=\abs{m}=2\) waveforms. 
This seems to suggest that metric \(\Delta {\cal{B}}\) is optimal to maximize the contribution of 
higher-order modes. We ascertain this observation in the following section.

\section{Signal-to-noise ratio calculations}
\label{sec:snrs}

Since we are using a discrete set of NR waveforms, and these scale trivially with mass, we have considered BBHs with total mass \(M=60\msun\), which is in agreement with the expected masses of compact binary populations in dense stellar environments~\cite{RoSa:2018}. This choice also implies that the merger of these BBHs will take place in advanced LIGO's optimal sensitivity frequency range, thereby maximizing the contribution of \((\ell, \, \abs{m})\) modes for SNR calculations. 

To quantify the SNR increase due to the inclusion of higher-order waveform modes, we use Eq.~\eqref{strain} for two cases, either all \((\ell, \, \abs{m})\) modes are included, or the \(\ell=\abs{m}=2\) is considered. We then consider the maxima \((\theta^{*},\,\phi^{*})\) of the \({\cal{A}}^{\left(\ell, \, \abs{m}\right)}\) and 
\({\cal{B}}^{\left(\ell, \, \abs{m}\right)}\) metrics, and construct their corresponding NR waveforms. Using these waveforms, we compute their corresponding SNRs at each sky location \((\alpha,\,\beta)\) and polarization angle \(\psi\), i.e., 

\begin{eqnarray}
\textrm{SNR}^{\left(\ell=\abs{m}=2\right)}&\rightarrow& \textrm{SNR} \left(\ell=\abs{m}=2;\, \theta^{*},\,\phi^{*};\,\alpha,\,\beta,\,\psi\right)\,,\nonumber\\
\textrm{SNR}^{(\ell,\,\abs{m})}&\rightarrow&\textrm{SNR} \left(\ell, \, \abs{m};\,  \theta^{*},\,\phi^{*};\,\alpha,\,\beta,\,\psi\right)\,.\nonumber
\end{eqnarray}

\noindent Finally, we subtract the aforementioned quantities and normalize them using the maximum value of  \(\textrm{SNR}^{\left(\ell=\abs{m}=2\right)}\) over the \((\alpha,\,\beta,\,\psi)\) parameter space, i.e., 

\begin{displaymath}
\textrm{SNR}^{\left(\ell=\abs{m}=2\right)}_{\textrm{max}} = \textrm{SNR}\left(\ell=\abs{m}=2;\, \theta^{*},\,\phi^{*};\,\hat{\alpha},\,\hat{\beta},\,\hat{\psi}\right)\,.
\end{displaymath}

\noindent Putting these quantities together leads to 

\begin{equation}
\Delta \textrm{SNR}= \frac{\textrm{SNR}^{(\ell,\,\abs{m})} - \textrm{SNR}^{\left(\ell=\abs{m}=2\right)} }{ \textrm{SNR}^{\left(\ell=\abs{m}=2\right)}_{\textrm{max}}}\,.
\label{snr_difference}
\end{equation}

\begin{figure*}
\centerline{
\includegraphics[width=.35\textwidth]{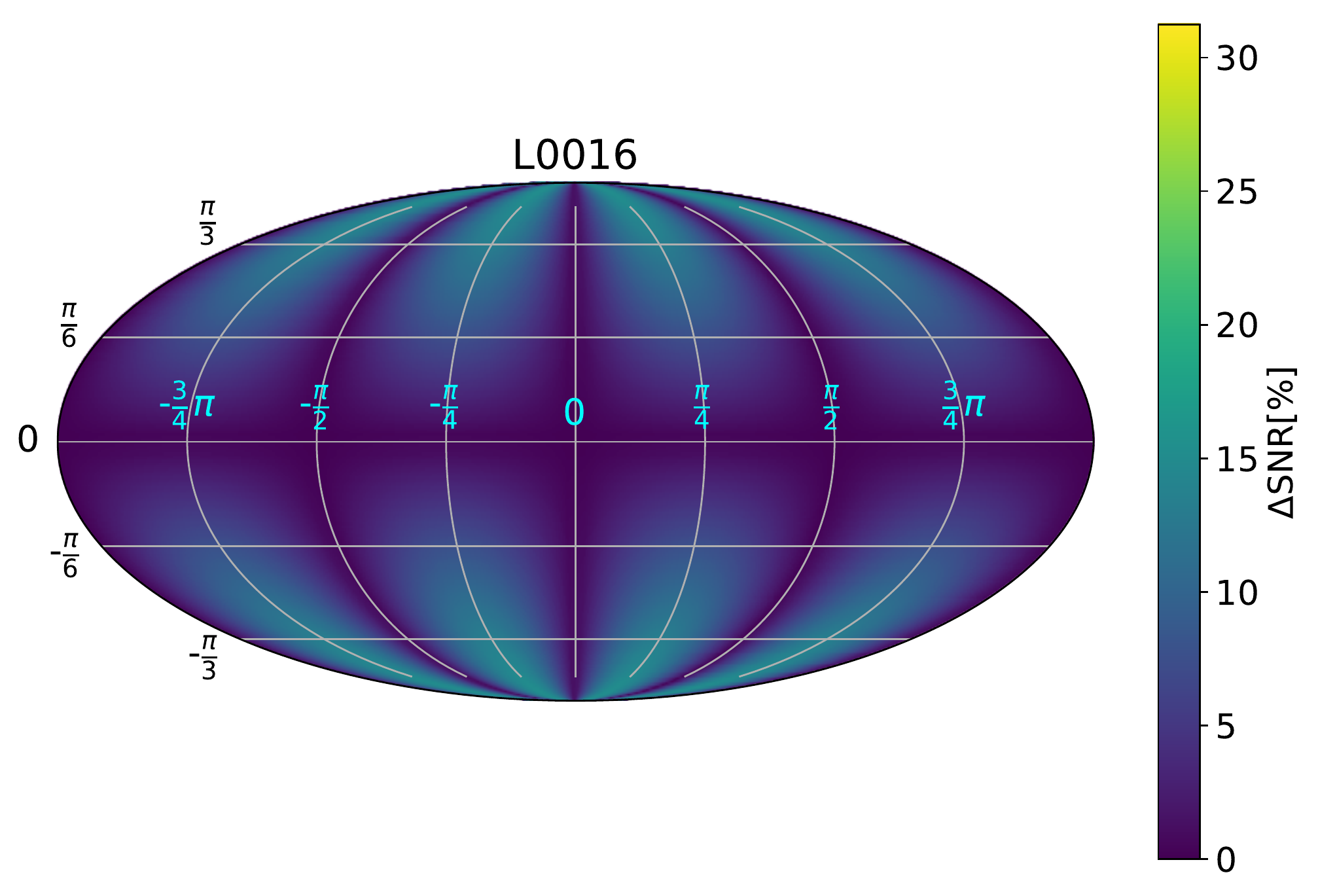}\hspace{8mm}
\includegraphics[width=.35\textwidth]{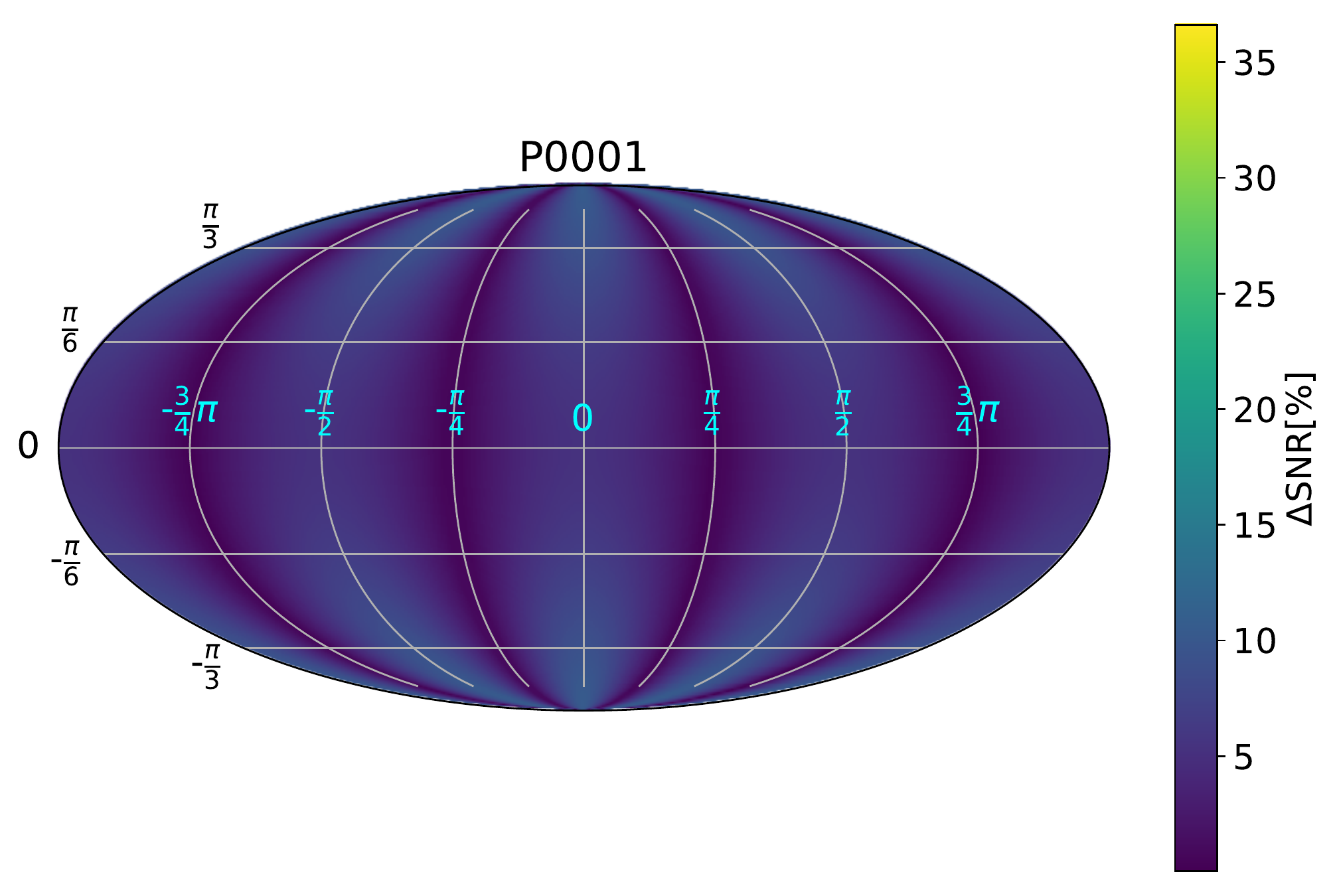}
}
\centerline{
\includegraphics[width=.35\textwidth]{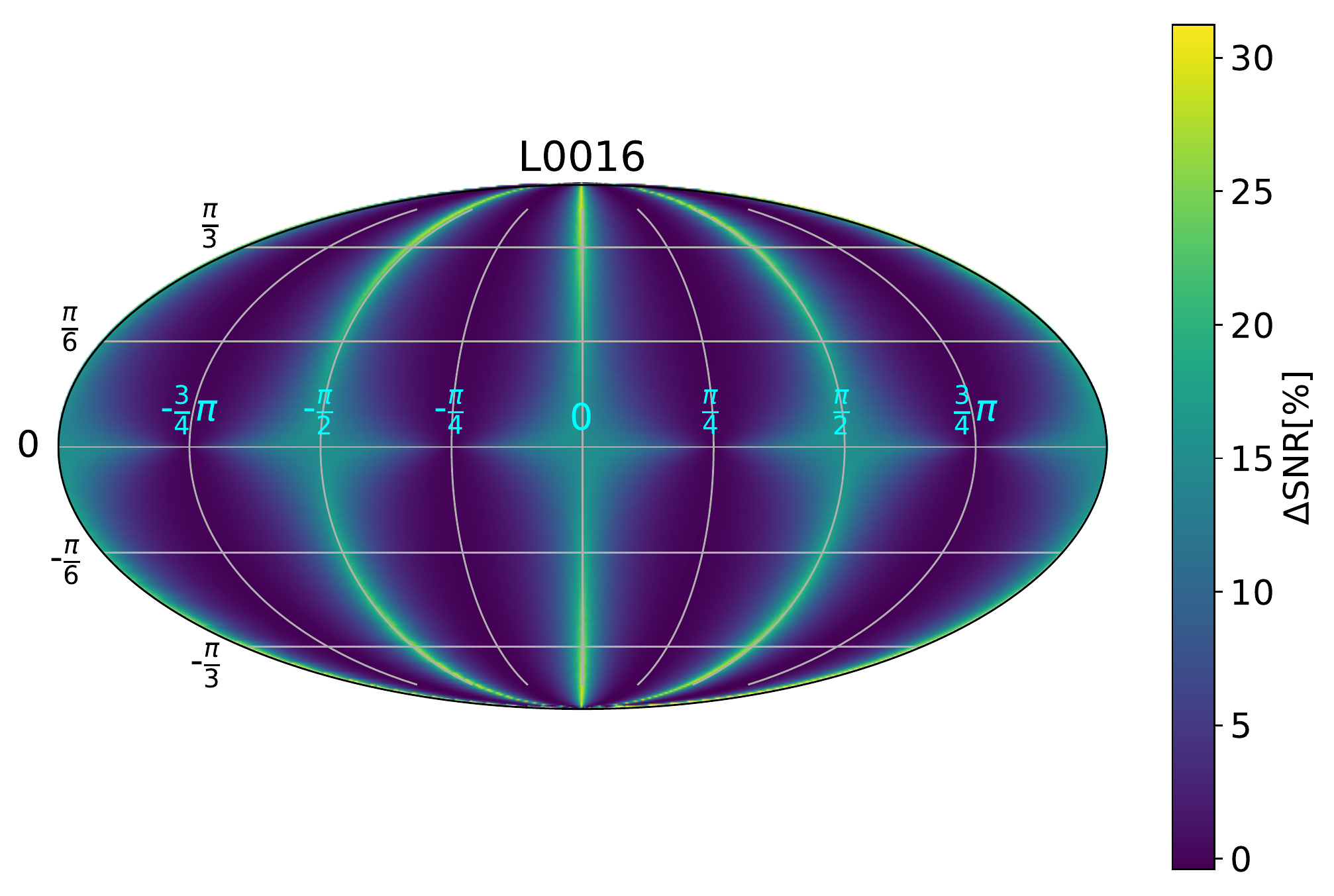}\hspace{8mm}
\includegraphics[width=.35\textwidth]{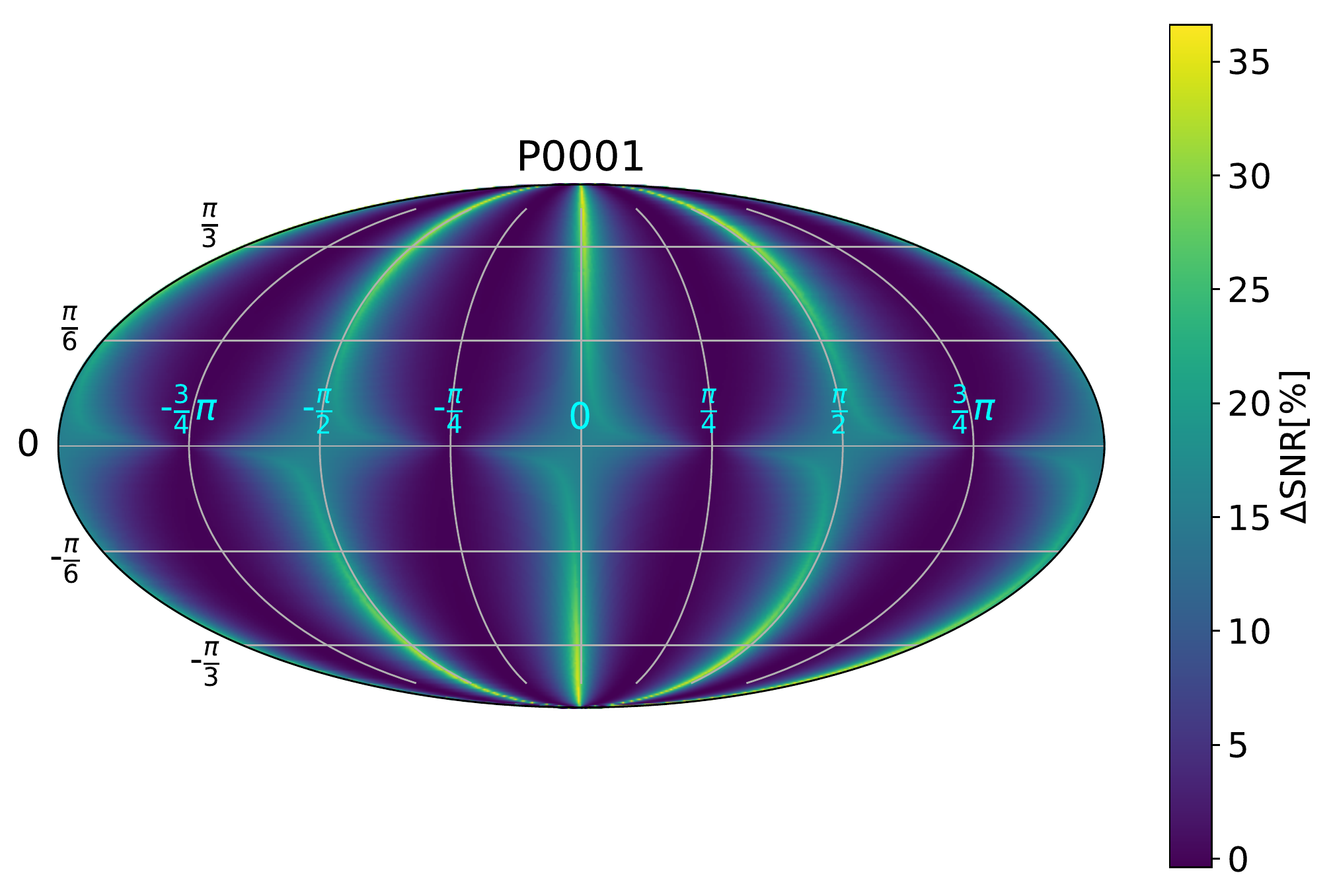}
}
\caption{Top panels: \(\Delta \textrm{SNR}\) distributions, see Eq.~\eqref{snr_difference}, using the maxima of metric \(\Delta {\cal{A}}^{\left(\ell, \, \abs{m}\right)}\). Bottom panels: as before, but now using the maxima of metric \(\Delta {\cal{B}}^{\left(\ell, \, \abs{m}\right)}\).  The  \(\Delta \textrm{SNR}\) distributions are presented as a function of the source's sky location \((\alpha,\,\beta\)) mapped into a Mollweide projection: \((\vartheta, \varphi) \rightarrow (\pi/2-\alpha, \beta-\pi)\). We have set \(\psi=\pi/4\) in these calculations.}
\label{comparison_snr}
\end{figure*}

\noindent Figure~\ref{comparison_snr} presents the \(\Delta \textrm{SNR}\) distributions for the  \(\Delta {\cal{A}}^{\left(\ell, \, \abs{m}\right)}\) and 
\(\Delta {\cal{B}}^{\left(\ell, \, \abs{m}\right)}\) metrics for two sample cases in which we have found a significant SNR increase. In practice, we have constructed two waveform families using the maxima of the \(\Delta {\cal{A}}\) and \(\Delta {\cal{B}}\) metrics, respectively, as shown in Figures~\ref{maximization_A} and~\ref{maximization}. These results show that the \((\theta^{*},\,\phi^{*})\) regions determined by metric \(\Delta {\cal{B}}^{\left(\ell, \, \abs{m}\right)}\) indeed lead to a more significant increase in SNR, as argued in the previous section. This is expected, since this metric was explicitly constructed to constrain the regions where the inclusion of higher-order modes would lead to the maximum integrated amplitude of the GWs. If we now consider BBHs whose masses are such that the merger takes place in the sensitive frequency band of GW detectors, then it follows that metric \(\Delta {\cal{B}}^{\left(\ell, \, \abs{m}\right)}\) is indeed the optimal one to obtain the maximum increase in SNR.

\begin{figure*}
\centerline{
\includegraphics[width=.34\textwidth]{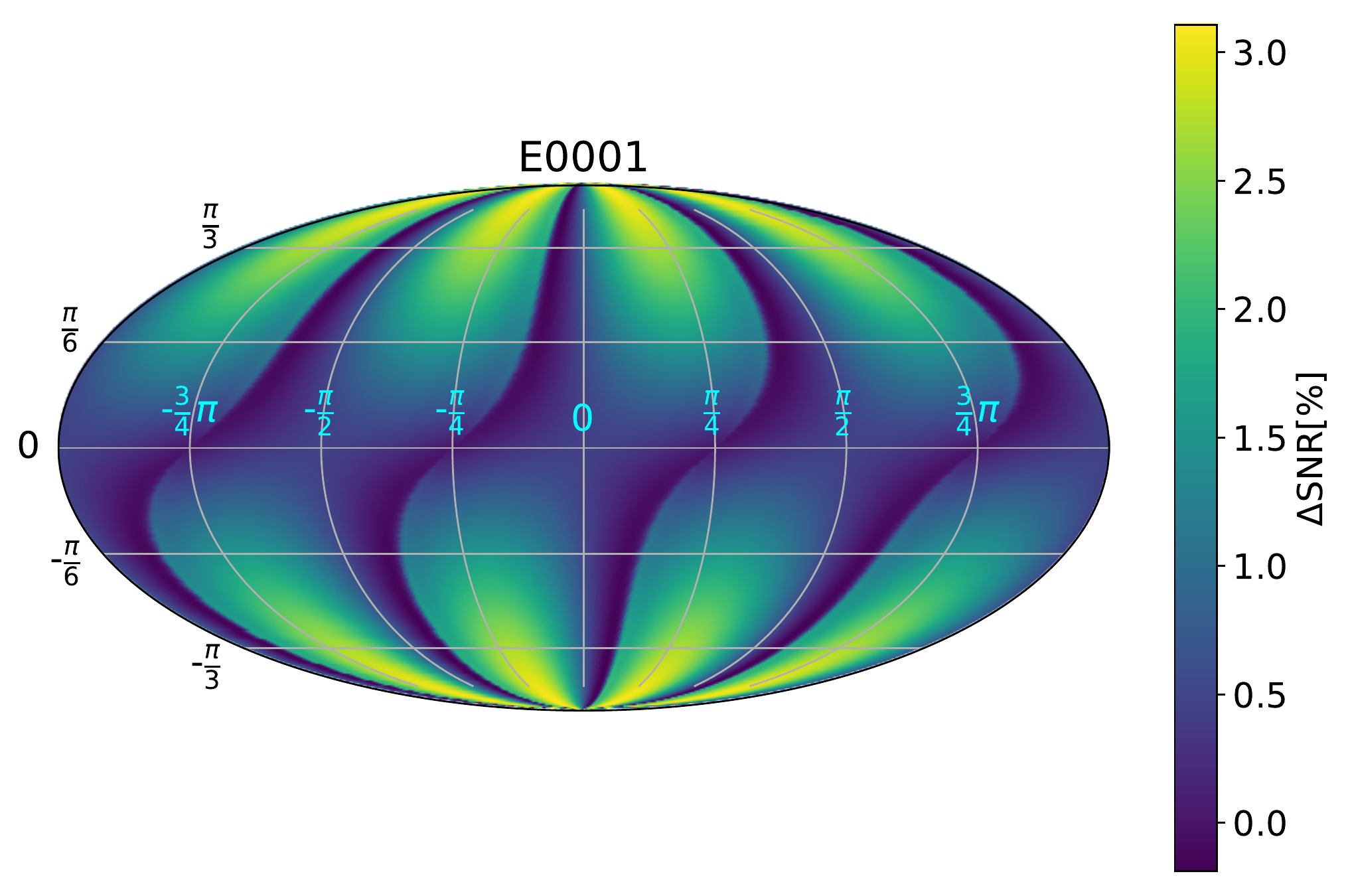}\hspace{0.1em}%
\includegraphics[width=.34\textwidth]{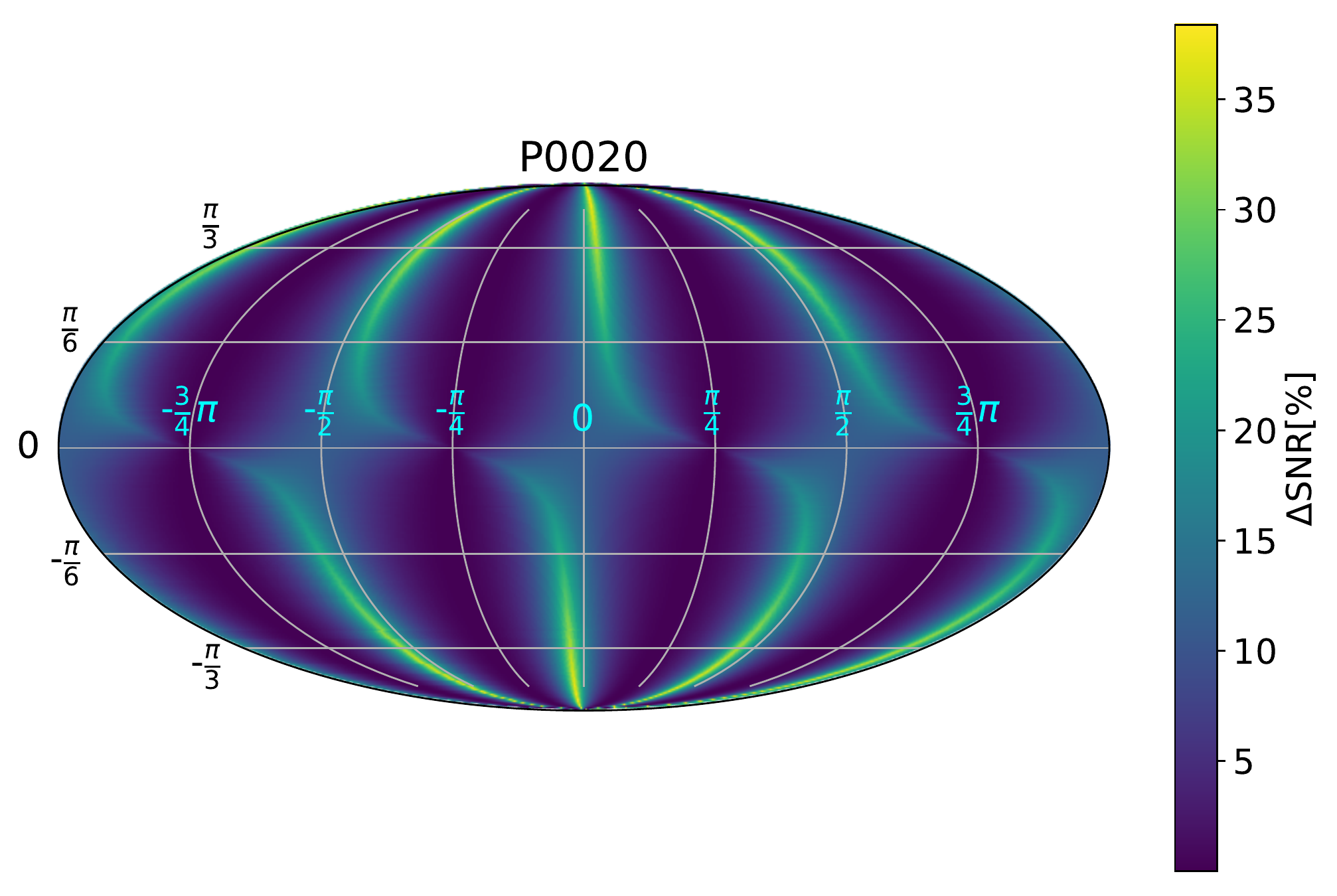}\hspace{0.1em}%
\includegraphics[width=.34\textwidth]{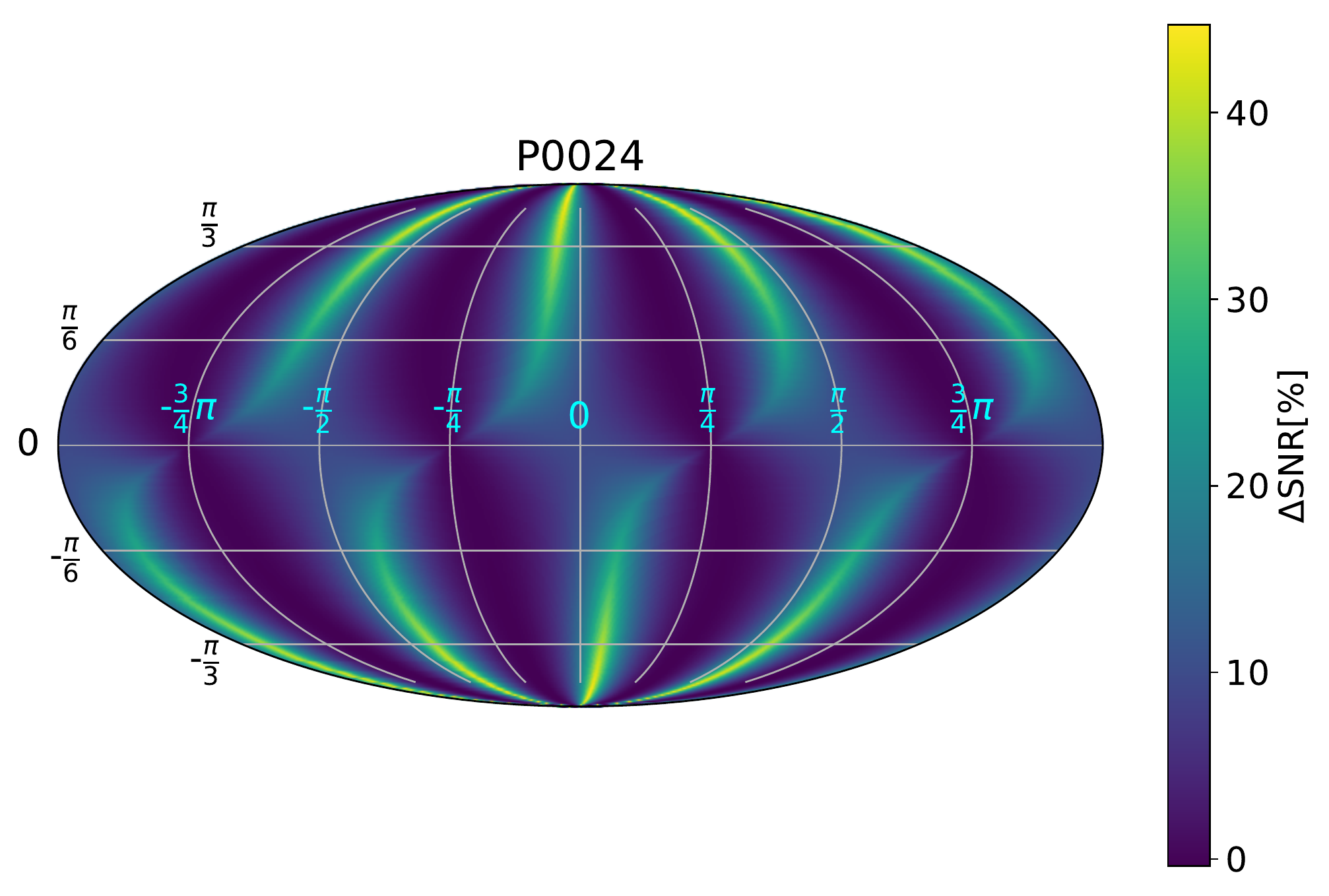}\hspace{0.1em}%
}
\centerline{
\includegraphics[width=.34\textwidth]{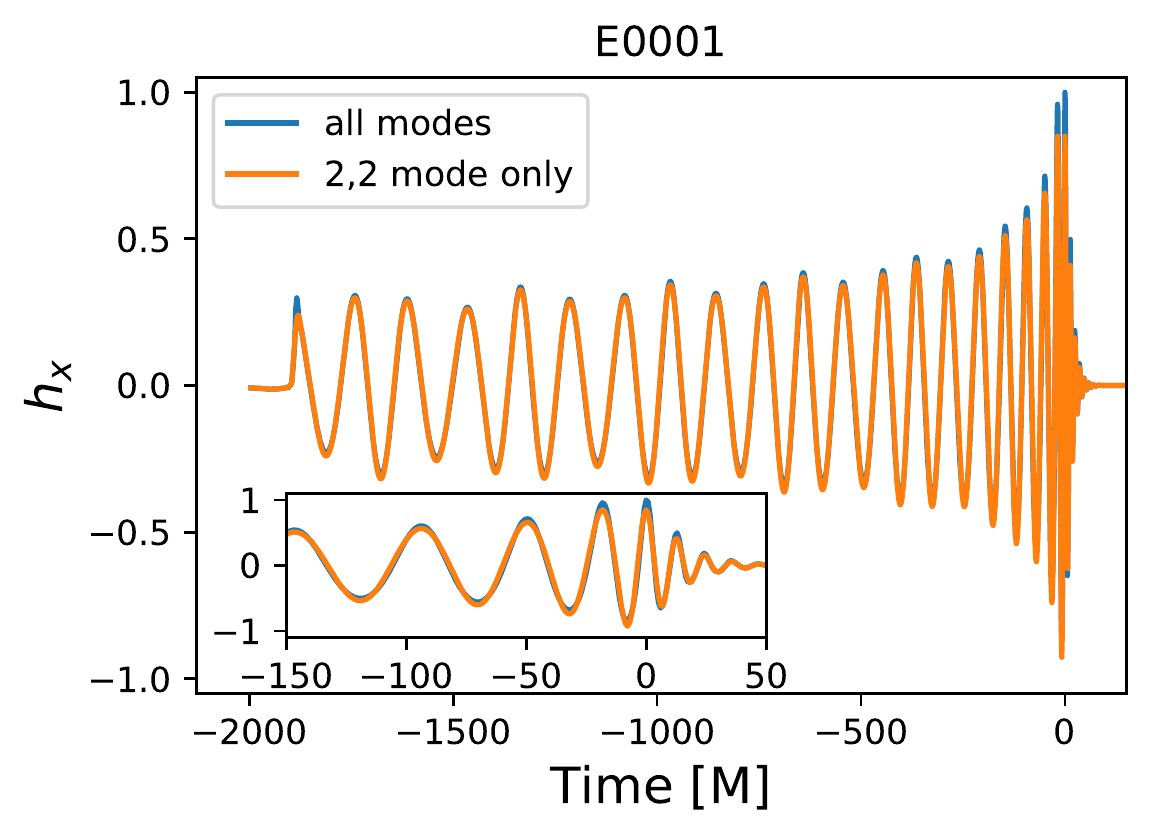}\hspace{.1em}%
\includegraphics[width=.34\textwidth]{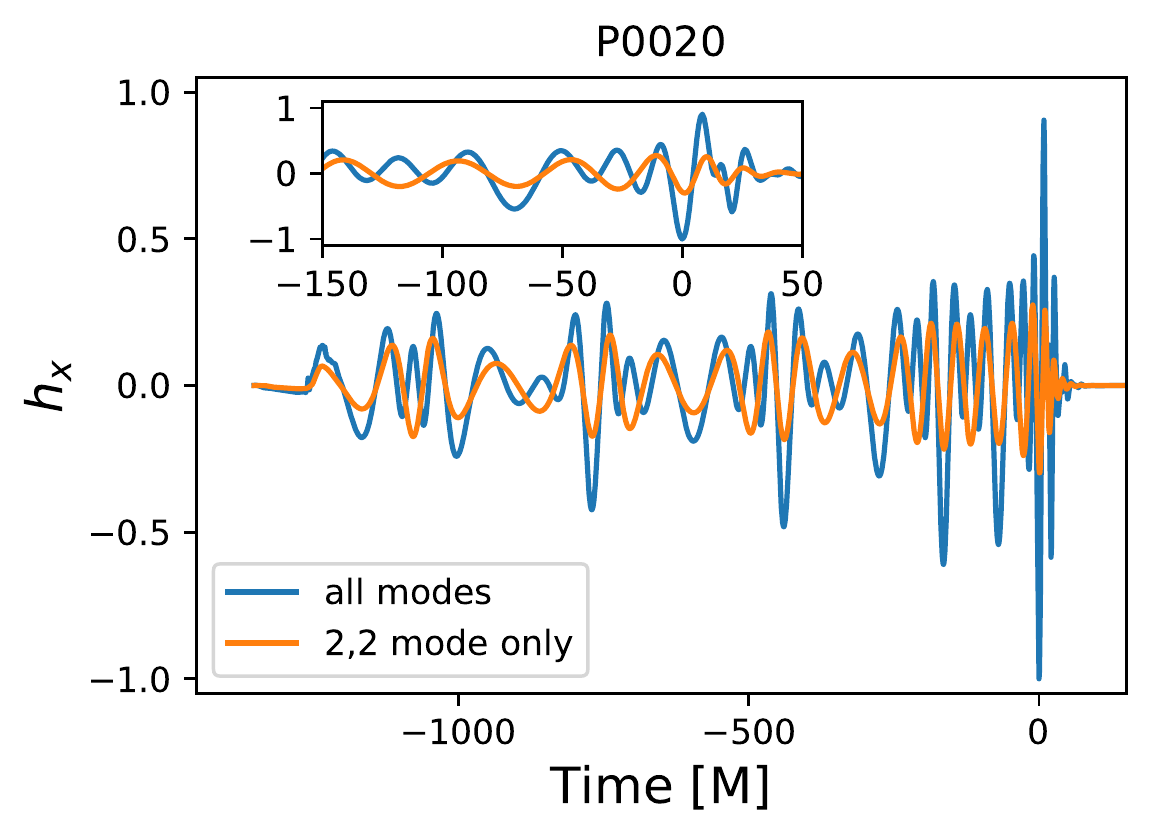}\hspace{0.1em}%
\includegraphics[width=.34\textwidth]{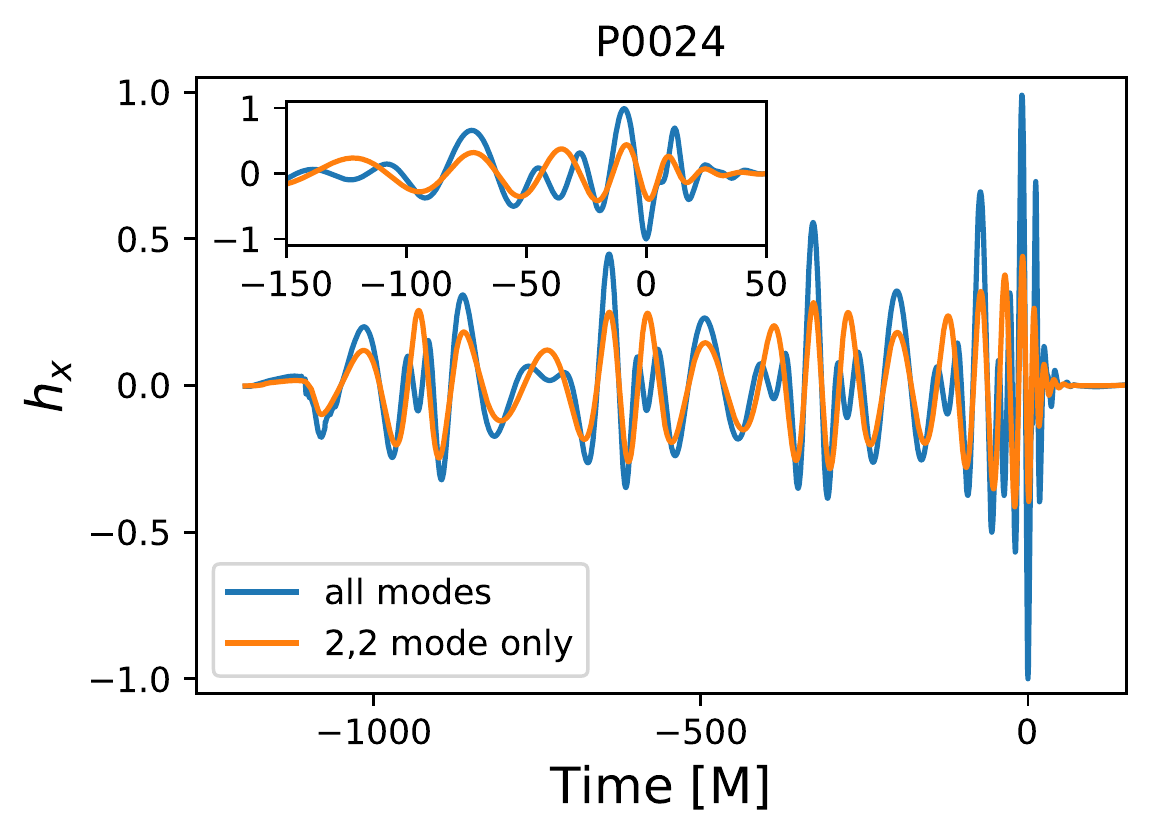}\hspace{0.1em}%
}
\caption{Top panels: \(\Delta \textrm{SNR}\) distributions, see Eq.~\eqref{snr_difference}, using the maxima of metric \({\cal{B}}^{\left(\ell, \, \abs{m}\right)}\). The SNR distributions are presented as a function of the source's sky location \((\alpha,\,\beta\)) mapped into a Mollweide projection: \((\vartheta, \varphi) \rightarrow (\pi/2-\alpha, \beta-\pi)\). We have set \(\psi=\pi/4\) in these calculations. Bottom panels: comparison between NR waveforms that include either all \((\ell, \, \abs{m})\) modes or the  \(\ell=\abs{m}=2\) mode only, using \((\theta,\,\phi)\) values that correspond to the maxima of metric \({\cal{B}}^{\left(\ell, \, \abs{m}\right)}\). Visualizations for these \(\Delta \textrm{SNR}\) distributions for a continuous range of \(\psi\) values can be found at~\cite{viz_homodes_dl}.}
\label{snrs}
\end{figure*}

Based on the aforementioned observations, Figure~\ref{snrs} presents SNR distributions for additional cases using metric \(\Delta {\cal{B}}^{\left(\ell, \, \abs{m}\right)}\). We also include the corresponding waveforms used for these calculations. While these results have been produced for a fixed polarization angle \(\psi\) value, we have also produced visualizations for \(\Delta \textrm{SNR}\) for a continuous \(\psi\) range. These are available at~\cite{viz_homodes_dl}. Please note that we have purposefully chosen not to average these results over the polarization angle for two reasons: (i) the in-depth analysis presented in~\cite{Schutz:2011} explicitly states that it is always better to explore in detail the impact of the polarization angle in the SNR of compact binary populations, since polarization-averaged results are only useful to study the gross properties of detection; (ii) we have discussed that \((h_+,\,h_{\times})\) in Eq.~\eqref{strain} differ significantly from each other, even at the leading-order post-Newtonian. Therefore, by explicitly computing the SNR distribution as a function of the source's sky location, \((\alpha,\,\beta\)), and the polarization angle, \(\psi\), we get the full picture of the interplay between all these components to the actual SNR distribution of a given BBH population.

\noindent The top-left panel in Figure~\ref{snrs} indicates that higher-order modes can be safely ignored for equal mass BBHs. However, the inclusion of \((\ell, \, \abs{m})\) modes becomes of paramount importance for asymmetric mass-ratio BBH mergers, as shown in the top-middle and top-right panels of Figure~\ref{snrs}. In these two panels, we notice that dark stripes in the sky maps correspond to regions of parameter space where \(\ell=\abs{m}=2\) NR waveforms have negligible SNRs. However, once we include higher-order modes, the SNR of  \((\ell, \, \abs{m})\) NR waveforms in these same regions is equivalent to 40\% and 45\%, respectively, of the SNR of an optimally oriented \(\ell=\abs{m}=2\) NR waveform. Furthermore, we learn from the bottom panels of Figure~\ref{snrs} that the inclusion of \((\ell, \, \abs{m})\) modes significantly modifies the ringdown evolution of \(\ell=\abs{m}=2\) NR waveforms. The finding is in line with studies that indicate the need to include \((\ell, \, \abs{m})\) modes for tests of general relativity using ringdown waveforms~\cite{BertiSes:2016PRL,YagiSCQG:2016,Pang:2018PRD,Berti:2018GReGr,BertYagi:2018GReGr}. 


\subsection{Interplay of mass-ratio and eccentricity for gravitational wave detection}
\label{supp2}

\begin{figure*}
\centerline{
\includegraphics[width=.34\textwidth]{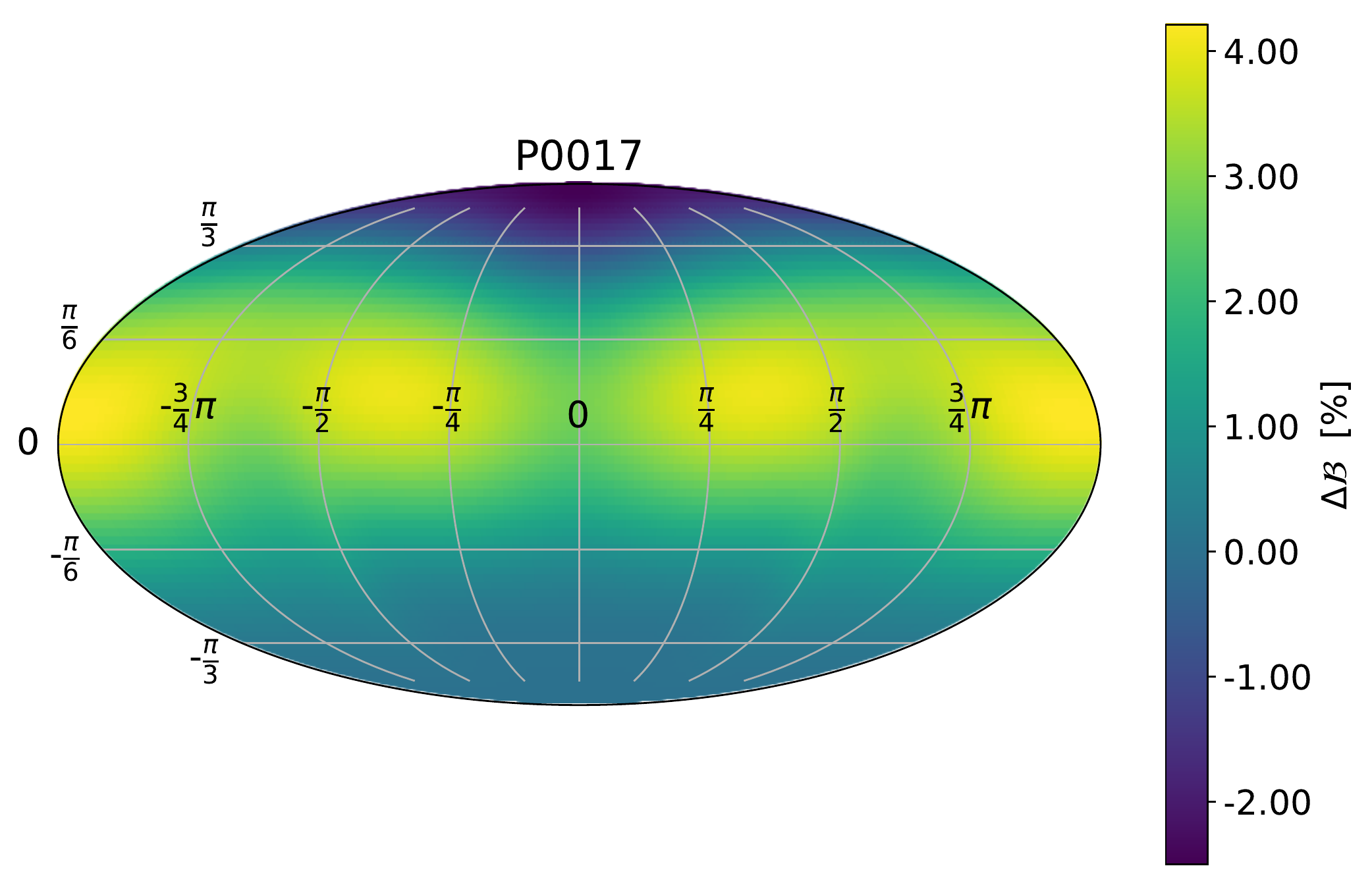}\hspace{8mm}%
\includegraphics[width=.34\textwidth]{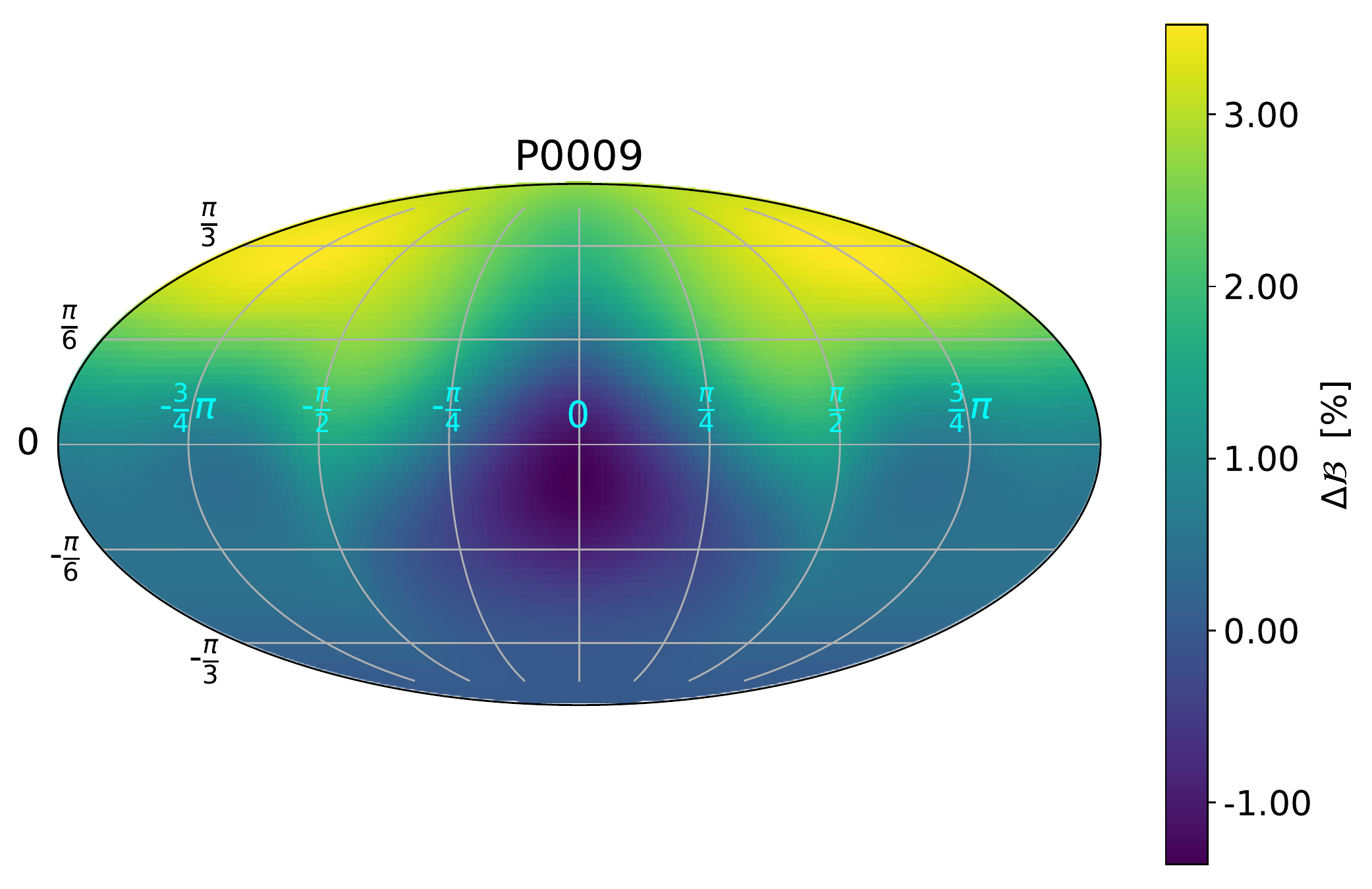}
}
\centerline{
\includegraphics[width=.34\textwidth]{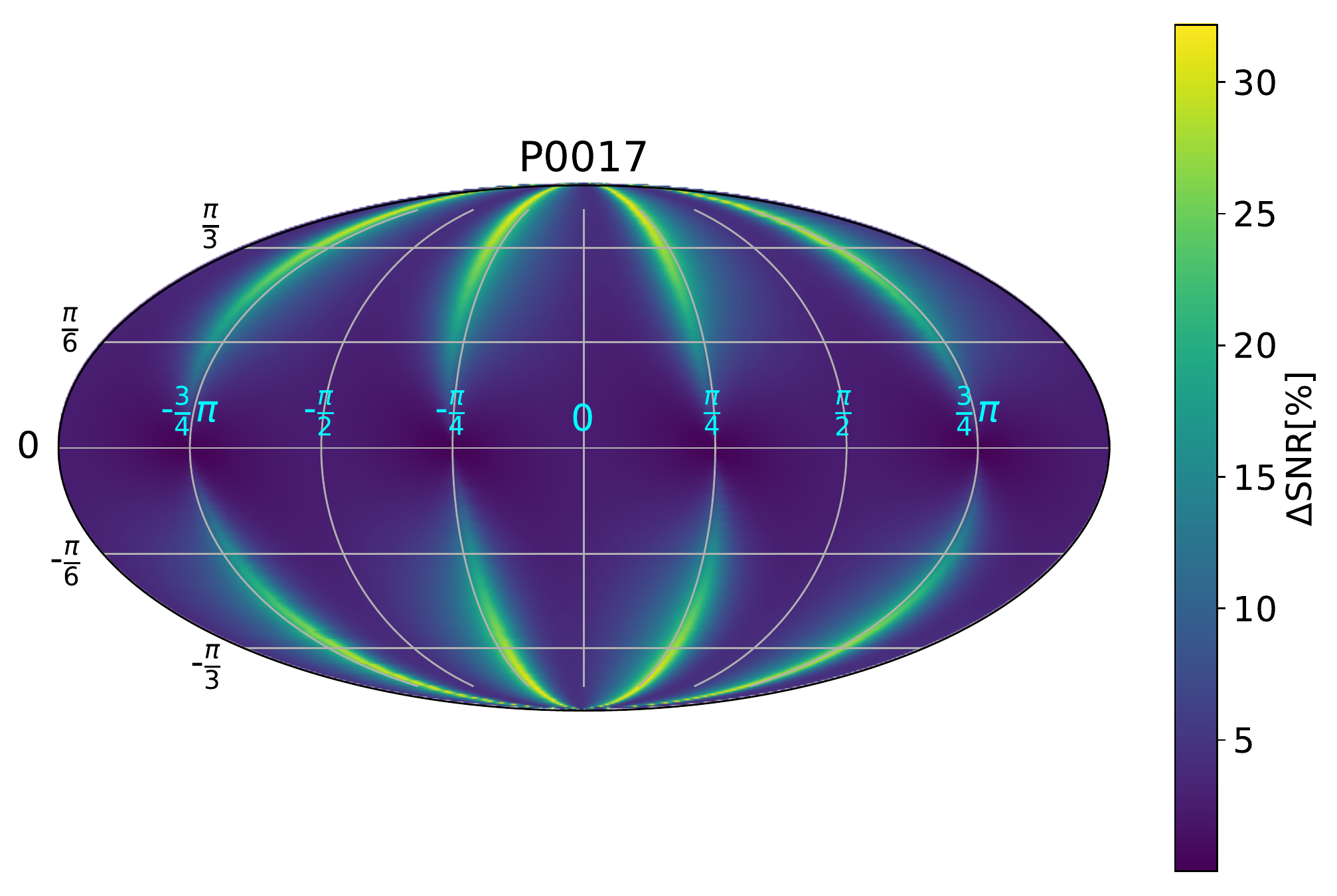}\hspace{8mm}%
\includegraphics[width=.34\textwidth]{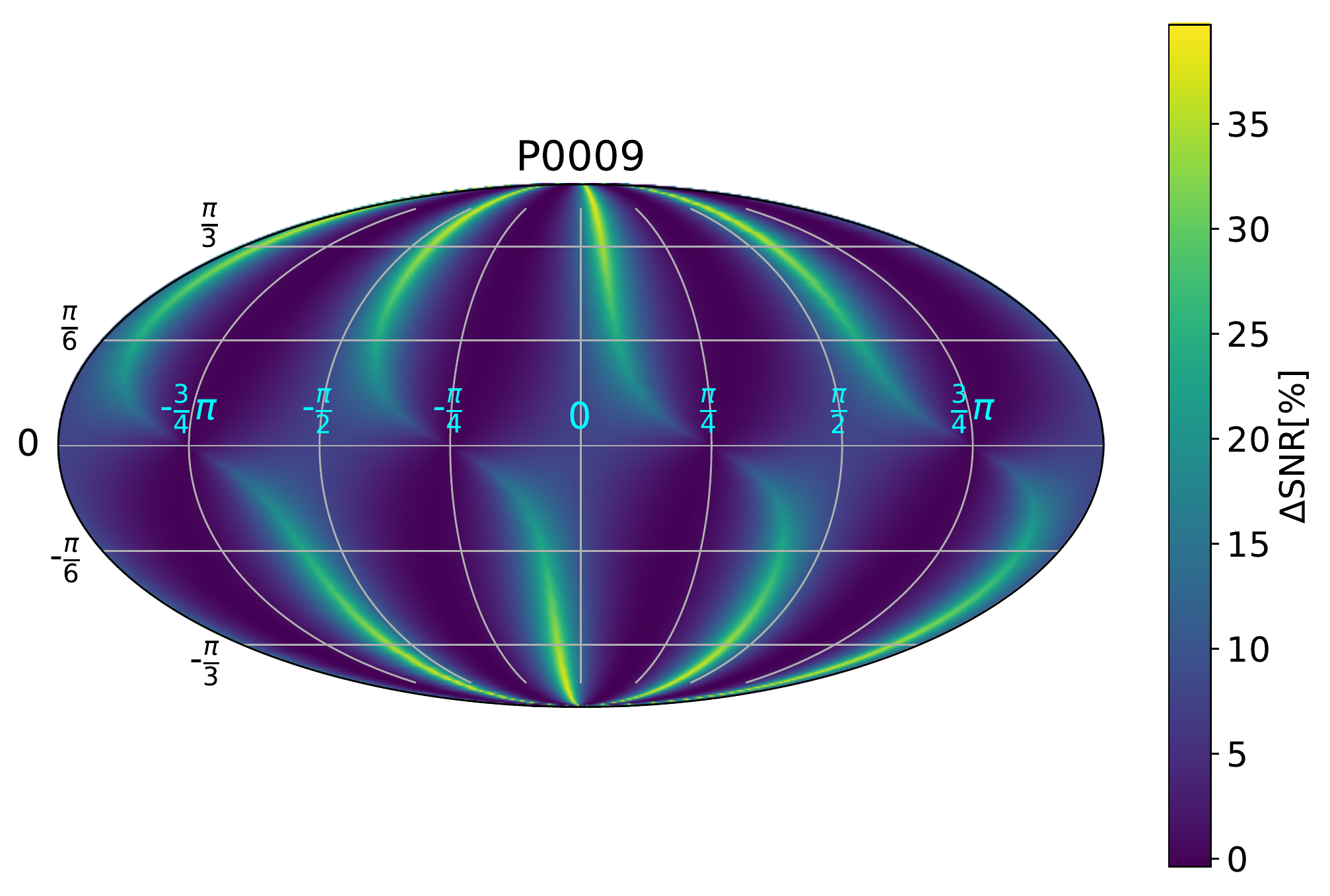}
}
\caption{Pair-wise comparisons of BBH systems (P0017,\, P0020) and (P0009,\, P0024).  (P0017,\, P0020) have mass-ratio \(q=8\), whereas (P0009,\, P0024) have mass-ratio \(q=10\). From each pair, P0017 and P0009 are the least eccentric. Top panels: \((\theta,\,\phi)\) regions that maximize the contribution of higher-order modes for GW detection using metric \(\Delta {\cal{B}}^{\left(\ell, \, \abs{m}\right)}\). Compare these results to the angular distributions for (P0020,\,P0024) in Figure~\ref{maximization}. Bottom panels: \(\Delta \textrm{SNR}\) distributions assuming \(\psi=\pi/4\). Compare these results to those presented in Figure~\ref{snrs}.}
\label{dis}
\end{figure*}

\noindent To get insights into the interplay that mass-ratio and eccentricity  have in the angular distributions that maximize the contribution of higher-order modes, the top panels of Figure~\ref{dis} present the \((\theta,\,\phi)\) angular distributions using metric \(\Delta {\cal{B}}^{\left(\ell, \, \abs{m}\right)}\), and the corresponding \(\Delta \textrm{SNR}\) distributions for (P0017,\,P0009). These results can be directly compared with those obtained for (P0020,\, P0024), presented in Figures~\ref{maximization} and Figure~\ref{snrs}. (P0017,\, P0020) have mass-ratio \(q=8\), whereas (P0009,\, P0024) have mass-ratio \(q=10\). From each pair, P0017 and P0009 are the least eccentric. These results indicate that while mass-ratio is the key parameter that drives the increase in SNR, 
eccentricity plays two important roles: (i) for a fixed-mass-ratio BBH population, larger eccentricities produce louder signals; (ii) eccentricity determines the inclination and sky regions where eccentric BBH mergers are optimally detected. These regions of parameter space are rather distinct for different eccentricity values.
    
\noindent This section concludes our studies on the importance of including higher-order modes for GW detection in terms of SNRs. It is worth commenting that a similar study in the context of quasi-circular, non-spinning BBH mergers was presented in~\cite{Prayush:2013a,Colin:2013} (see also~\cite{Pekowski:2013,Bustillo:2015qty,Varma:2016dnf} for spinning, quasi-circular BBH mergers). In that study, it was found that the inclusion of higher-order modes could increase SNR calculations by as much as 8\%. Having used the same advanced LIGO PSD for these studies, we find that higher-order modes play a much more significant role in the dynamics of eccentric BBH mergers.


\section{Impact of \((\ell, \, \abs{m})\) modes on the range of detectability for eccentric binary black hole mergers}
\label{range}

In this section we quantify the increase in detection range by a LIGO-type detector when we include higher-oder waveform modes. To do so, we compute the range within which a GW source can be observed using the definition~\cite{Schutz:2011}

\begin{equation}
D_{\textrm{obs}}=\frac{\left(h,\,h\right)}{\hat{\rho}}\,.
\label{eq_distance}
\end{equation}

\noindent \(\left(h,\,h\right)\) is given by Equation~\eqref{snr_white}, and \(\hat{\rho}=8\) is taken as a typical SNR detection threshold~\cite{Prayush:2013a}. Using this relation, we have computed \(D_{\textrm{obs}}\) for NR waveforms that include either all \((\ell, \, \abs{m})\) modes or just the \(\ell=\abs{m}=2\) mode, namely \(D^{(\ell, \, \abs{m})}_{\textrm{obs}}\) and \(D^{(\ell=\abs{m}=2)}_{\textrm{obs}}\), respectively. We define \(D^{(\ell=\abs{m}=2)}_{\textrm{obs,\, max}}\) as the maximum value of the \(D^{(\ell=\abs{m}=2)}_{\textrm{obs}}\) distribution across the sky \((\alpha,\,\beta)\). Under these conventions, we evaluate the increase in detection range due to the inclusion of higher-order modes using the metric  

\begin{equation}
\Delta D_{\textrm{obs}}(\alpha,\,\beta) = \frac{D^{(\ell, \, \abs{m})}_{\textrm{obs}}(\alpha,\,\beta) - D^{(\ell=\abs{m}=2)}_{\textrm{obs}}(\alpha,\,\beta)}{D^{(\ell=\abs{m}=2)}_{\textrm{obs, max}}}\,.
\label{difference}
\end{equation}

\begin{figure*}
\centerline{
\includegraphics[width=.34\textwidth]{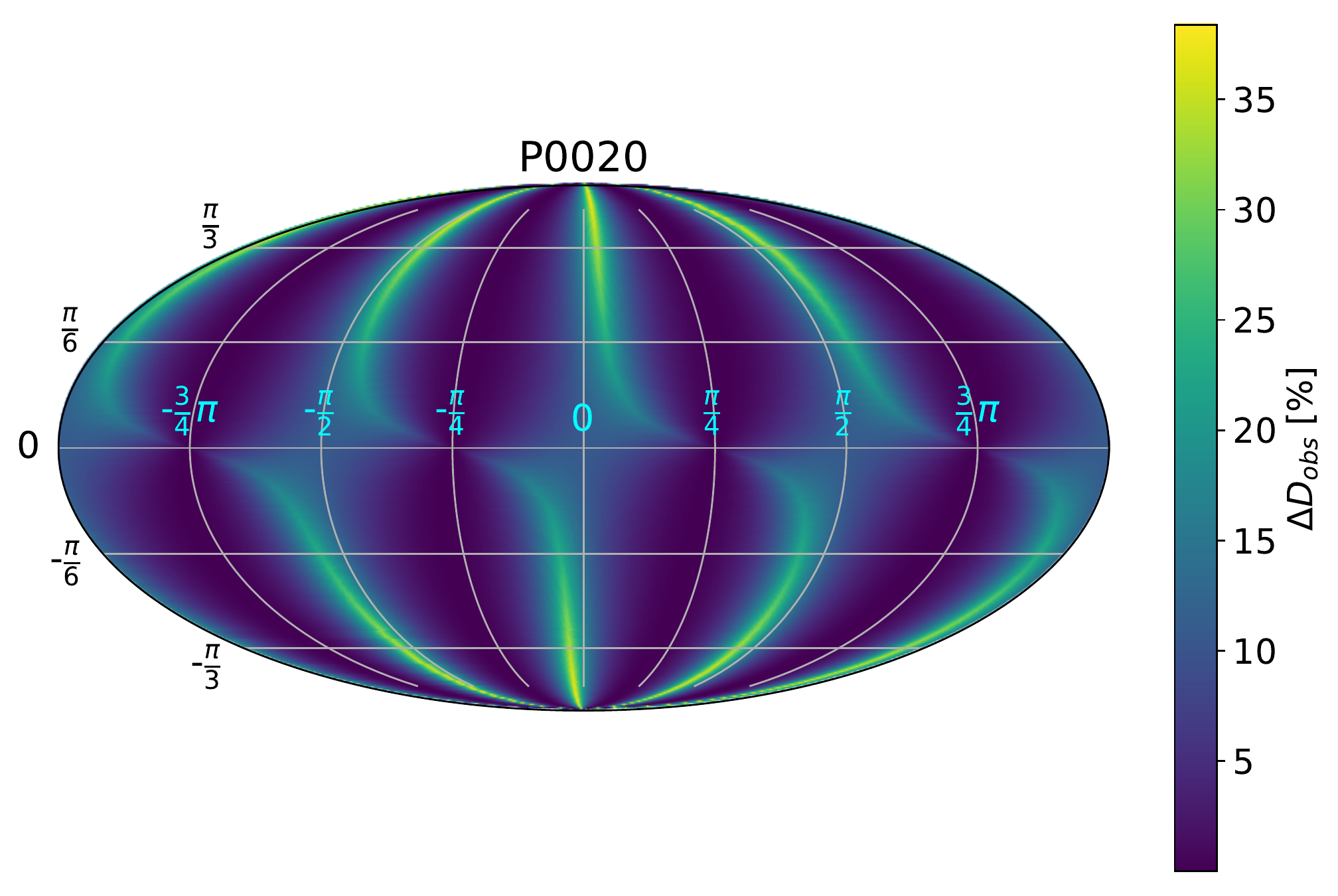}\hspace{8mm}%
\includegraphics[width=.34\textwidth]{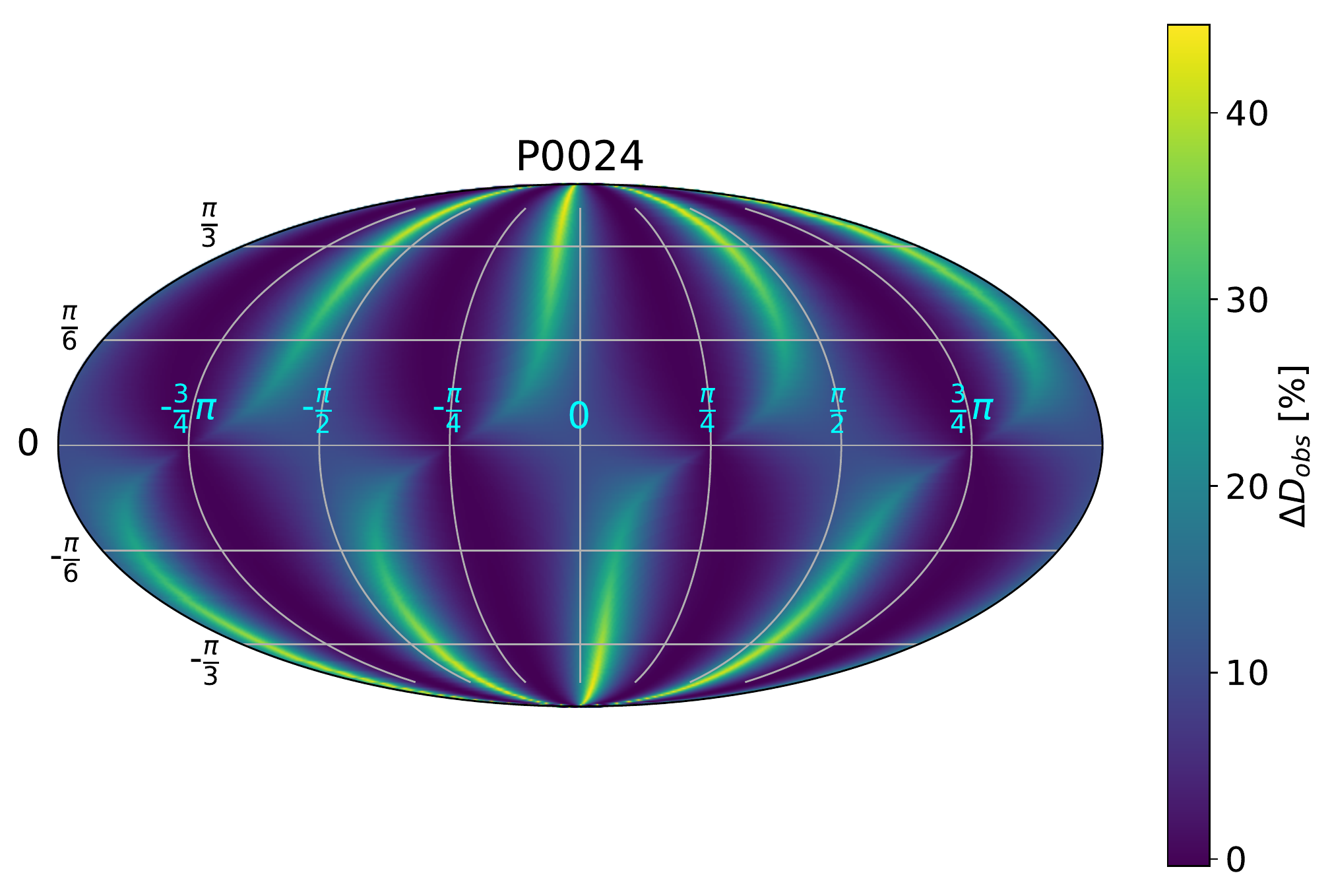}
}
\caption{Increase in detection range due to the inclusion of higher-order waveform multipoles for the eccentric binary black hole mergers described by the NR waveforms P0020 and P0024. These panels were produced setting \(\psi=\pi/4\). However, we have also produced visualizations that show the increase in \(D_{\textrm{obs}}\) using a continuous range of \(\psi\) values. These are available at~\cite{viz_homodes_dl}.}
\label{distance}
\end{figure*}

\noindent In Figure~\ref{distance} we present results for the two cases in which the inclusion of \((\ell, \, \abs{m})\) modes leads to the maximum increase in SNR with respect to \(\ell=\abs{m}=2\) NR waveforms, i.e., P0020 and P0024. Using these systems, we set the total mass of the binary \(M=60\Msun\), and the distance to the source to 500Mpc. As before, we use LIGO's ZDHP configuration~\cite{ZDHP:2010} for these calculations. 

The panels in Figure~\ref{distance} indicate that the inclusion of \((\ell, \, \abs{m})\) modes boosts the detection range of these types of GW sources by a similar amount as the effective increase in SNR that we report in Figure~\ref{snrs}. This is expected since the mathematical formalism used to estimate the increase in SNR and detection range when higher-order modes are included (see Eqs.~\eqref{snr_difference} and~\eqref{difference}) should produce the same answer. Note that while the panels in Figure~\ref{distance} are produced using \(\psi=\pi/4\), we have also produced visualizations in which we show the increase in detection range for a continuous range of \(\psi\) values. These visualizations are available at~\cite{viz_homodes_dl}.

Having quantified the importance of higher-order multipole waveforms in terms of SNR calculations, and the detection range of eccentric BBH mergers, in the following section we explore the detectability of these types of GW signals. At present, the most sensitive algorithms for GW detection are based on implementations of a matched-filter, and as discussed in~\cite{Sergey:2016} no matched-filtering algorithm has been presented in the literature that is tailored for the detection of \(\ell=\abs{m}=2\) eccentric waveforms~\cite{Sergey:2016}, let alone for higher-order multipole waveforms of eccentric BBH mergers. Furthermore, as mentioned before, template agnostic pipelines have been utilized to search for eccentric BBH mergers~\cite{Tiwari:2016,Sergey:2016}. However, these algorithms are optimal for the detection of burst-like GW signals, and may miss \({\cal{O}}(\textrm{second-long})\) signals with low SNRs~\cite{secondBBH:2016}. To circumvent this problem, in the following section we introduce a deep learning that is adequate for the detection of higher-order multipole waveforms from eccentric BBH mergers both in simulated and real advanced LIGO noise.


\section{Detection of eccentric binary black hole mergers with deep learning algorithms}
\label{ecc_det}

The application of deep learning across science domains is a booming enterprise. In the context of GW astrophysics, deep learning was introduced for the detection and characterization of GWs in simulated and real advanced LIGO noise in~\cite{geodf:2017a,geodf:2017b}. These results sparked the interest of the GW community, and have led to a number of different studies, including the modeling, detection and denoising of BBH mergers~\cite{wei:2019W,shen_icasp:2019,AlvinC:2018,Fan:2018,Gonza:2018,2018GN,Fuji:2018,LiYu:2017,hshen:2017,Nakano:2018,dglitcha:2017,dgNIPS,DL_Workshop:2018}. Recent applications of deep learning, developed by authors of this paper, have demonstrated that neural network models at scale can be used for real-time GW parameter estimation, and to reconstruct the astrophysical parameters of the remnants of spinning BHs that evolve on quasi-circular orbits~\cite{shen_dlscale:2019}. These deep learning algorithms have been used to estimate the astrophysical parameters of all the BBH mergers detected to date by the advanced LIGO and Virgo detectors, demonstrating that deep learning results are consistent with Bayesian analyses~\cite{o1o2catalog}. To further advance this science program, in this section we demonstrate that deep learning can detect and characterize higher-order waveform multipoles signals emitted by eccentric BBH mergers. 

Given that the topology of GWs that include higher-order waveform modes is much more complex than the NR waveforms we used in our previous studies~\cite{geodf:2017a,geodf:2017b}, we have significantly improved our deep learning algorithms by implementing the following features

\begin{cititemize2}
\item We have designed and trained our neural network models using \texttt{TensorFlow}~\cite{abadi2016tensorflow}. We have then combined our neural network models with \texttt{Horovod}~\cite{2018arXiv180205799S} to train our algorithms at scale. This approach has reduced the training stage from ten hours to just thirty minutes. We have tested this new approach in the Blue Waters supercomputer~\cite{Kramer2015,Bode2013} and the NCSA Innovative Systems Lab~\cite{ISLVlad} using NVDIA Tesla V100 GPUs.
\item  We have designed a new curriculum learning algorithm with an exponential decaying scheme, which proved critical to enhance the robustness and sensitivity of our neural network models for low SNR signals. Curriculum learning is used to train the neural network models to identify GWs over a broad SNR range, which is known as scale invariance. To do so, we start by exposing the neural network to GWs that have large SNRs. We then gradually decrease the SNR until the GW is completely embedded in noise, and has negligible SNR. This approach is critical to ensure that the deep learning algorithm is as sensitive as possible for low SNR signals, while also performing optimally for high SNR GWs. We use different realizations of noise in every iteration to prevent the neural network from memorizing noise, and to enhance its robustness and resilience when used in actual detection scenarios. 
\item We have also identified optimal hyperparameter values for batch size (number of waveforms processed before the neural network model is updated), learning rates and buffer size by running hundreds of parameter sweeps on the Blue Waters supercomputer. 
\item We have also ensured that our neural network models identify GWs anywhere in the data stream of GW detectors. Since we scan large batches of data using a 0.2 second time sliding window, we have trained our neural network models in such a way that a GW signal can be identified anywhere within a 0.2 second window. To make sure that the network does not memorize noise through this training procedure, we use different realizations of noise for each time-shift we apply to a given GW signal during the training stage. This approach ensures that our neural network models are time-invariant.
\item To increase the robustness and resilience of our neural network models to glitches (noise anomalies that obscure or mimic true GW signals), we also trained them using data sets of simulated sine-Gaussian glitches that span a broad range of frequencies, amplitudes, peak positions, and widths that are present in real advanced LIGO data~\cite{jade1:2016}. With this approach, we make our models resilient to glitches that are already present in the real advanced LIGO noise used for training and testing, and endow our neural network classifiers with the capability to tell apart glitches from true signals.
\end{cititemize2}

\subsection{Data curation}

We used the NR Surrogate model presented in~\cite{blackman:2015} to produce a dataset of quasi-circular, non-spinning BBH mergers to train our neural network model. Each waveform describes the last second of evolution of a given BBH merger, including the inspiral-merger-ringdown phases. We have used a sample rate of 8192Hz to ensure that the ringdown of low mass BBH systems is accurately sampled. A gallery of waveforms used for these analyses are presented in Figures~\ref{waveforms} and~\ref{waveforms_simulated}. We have considered BBHs with component masses \(5\Msun\leq m_{\{1,\,2\}} \leq 100\Msun\), such that the mass-ratio \(q\leq10\). The training dataset samples this range in steps of \(\Msun\), whereas the testing dataset uses intermediate masses in this range. 

We have trained and validated our neural network model with different types of noise: (i) simulated Gaussian noise; (ii) noise from the \href{hhttps://www.gw-openscience.org/about/}{Gravitational Wave Open Science Center}~\cite{losc} from the Hanford (H1) and Livingston (L1) detectors; and (iii) a mix of raw H1, L1 noise and simulated Gaussian noise. We tested our neural networks using only H1 and L1 noise from advanced LIGO's first observing run that was not used for training and validation. 

Our neural networks have been designed to process continuous data streams from a three-detector network H1-L1-V1. To train our neural network models, we prepare one second long data segments that contain either just noise or noise plus a GW signal. Once fully trained, we measured the false alarm rate of our neural network classifier by applying it on new inputs containing only real LIGO noise without any signals and counting how often false alarms were triggered. Using this procedure we found that our network triggered one false alarm every three months, i.e., one second of noise is misclassified as a signal once in every three months of searched data, which is adequate for an on-line search, the target use of these neural network models. As we mentioned in previous analyses~\cite{geodf:2017a,geodf:2017b}, at present these neural network models do not provide the significance of a given detection. This key component is in earnest development and will be presented in future work. 

\subsection{Simulated advanced LIGO noise}

\begin{figure*}
  \includegraphics[width=0.49\textwidth]{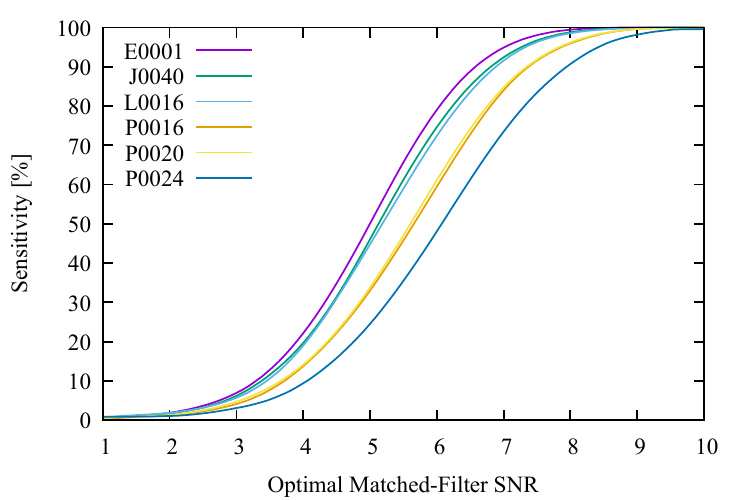}
  \includegraphics[width=0.49\textwidth]{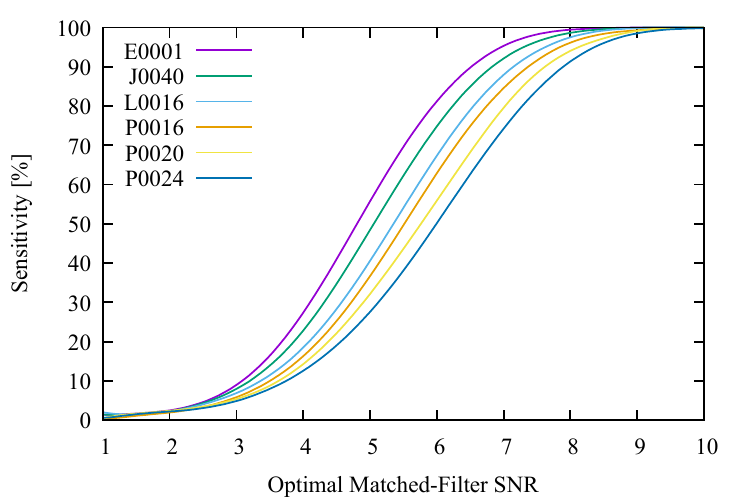}
   \caption{Left panel: Sensitivity of detecting higher-order waveform multipole signals injected in simulated LIGO noise. Right panel: as in left panel, but now injecting higher-order waveform multipole signals in real advanced LIGO noise. The binary black hole systems used in this study describe a broad range of eccentricities and mass-ratios to quantify the accuracy of our neural network models to extract waveforms in which higher-order modes have a marginal impact (comparable mass-ratios and moderate values of eccentricity) to configurations (high mass-ratio and large eccentricities) in which the injected signals are very distinct to our training dataset. In both scenarios, our neural network models reach 100\% Sensitivity to identify higher-order waveform multipole signals with \(\textrm{SNR}\geq 10\). The false positive rate of our neural network models is }
 \label{sensitivity}
 \end{figure*}

To quantify the probability of detecting a true signal (sensitivity) with which our deep learning algorithms can extract eccentric, higher-order waveform multipole signals as a function of SNR, we have selected waveform signals that represent a broad set of scenarios that describe cases in which higher-order modes have a marginal impact in the waveform signal (comparable mass-ratios and moderate values of eccentricity)  to challenging configurations (high mass-ratio and large eccentricities) in which the waveform signals are very distinct to our training dataset. All these waveforms have been constructed using the \((\theta, \,\phi)\) configurations that correspond to the maxima of the metric \(\Delta {\cal{B}}\), see Equation~\ref{area}. We have then whitened the signals and the Gaussian noise using LIGO's ZDHP configuration~\cite{ZDHP:2010}. Finally, we embed the whitened signals in whitened noise, and train the network using a broad range of SNRs using curriculum learning, as described in the previous section. The left panel of Figure~\ref{sensitivity} shows that independently of the \((q,\,e)\) of the BBHs, our deep learning algorithm achieves 100\% sensitivity for signals with \(\textrm{SNR}\geq10\). It is remarkable that this result is similar in the context of quasi-circular BBH signals~\cite{geodf:2017a}.  

\subsection{Real advanced LIGO noise}

For these studies we used open source real advanced LIGO noise available at the Gravitational Wave Open Science Center~\cite{losc}. For training we used data provided for the events GW170104, GW170608 and GW170729. For validation we used data for the events GW170814 and GW170818. Finally, for testing purposes we used advanced LIGO data of GW170817.

The right panel of Figure~\ref{sensitivity} indicates that our neural network model attains 100\% sensitivity to detect GWs with\(\textrm{SNR}\geq 10\). It is worth mentioning that, according to the study presented in~\cite{Nitz2018CQG}, the BBH systems we have used for this study would be affected by noise anomalies, thereby hurting the sensitivity of any algorithm used to search for GWs emitted by high mass BBH mergers. However, our results show that the performance of our deep learning algorithms is just marginally affected by the presence of noise anomalies in real advanced LIGO noise, as shown in the right panel of Figure~\ref{sensitivity}. This is expected, because we have exposed our neural network models to a broad range of scenarios that already include these types of noise anomalies, thereby making it resilient to them. We reported similar findings when recovering \(\ell=\abs{m}=2\) NR waveforms embedded in real advanced LIGO noise that describe non-spinning BHs both on quasi-circular and eccentric orbits in~\cite{geodf:2017b}.

\subsection{Parameter estimation}
\begin{figure*}[!htp]
\centerline{
\includegraphics[width=0.5\textwidth]{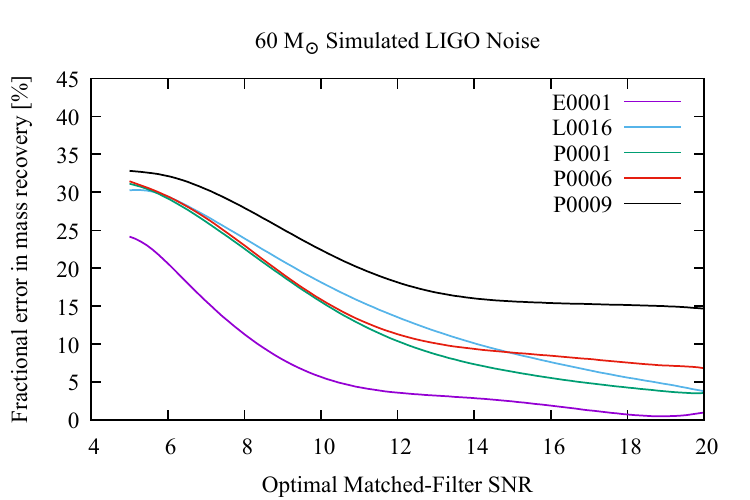}
\includegraphics[width=0.5\textwidth]{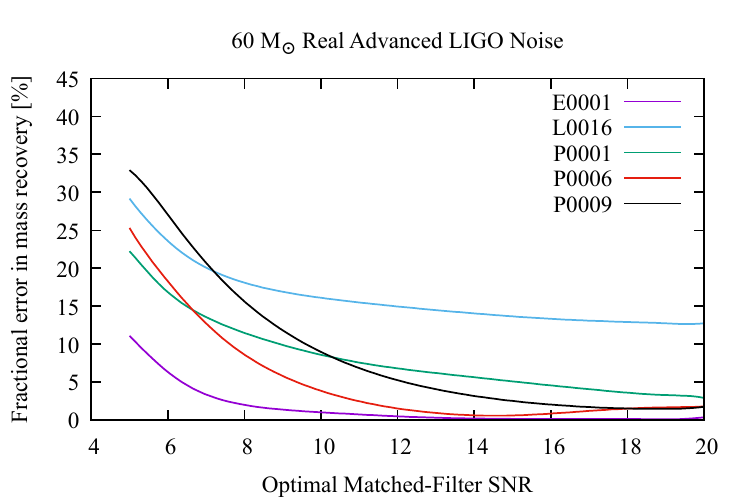}
  }
\centerline{  
 \includegraphics[width=0.5\textwidth]{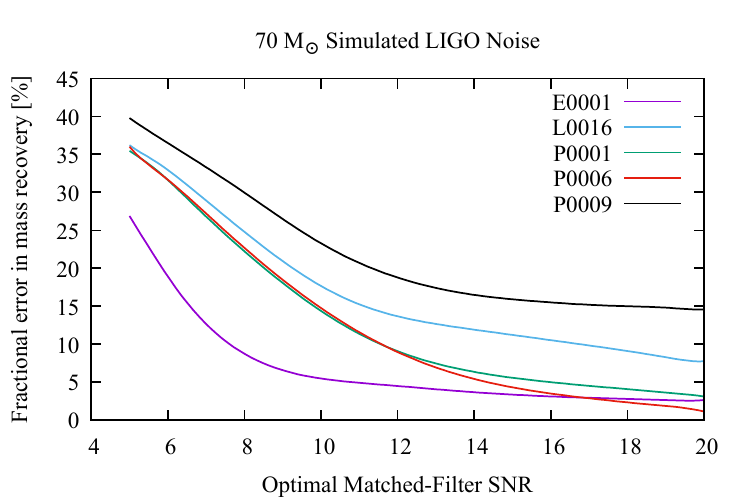}
  \includegraphics[width=0.5\textwidth]{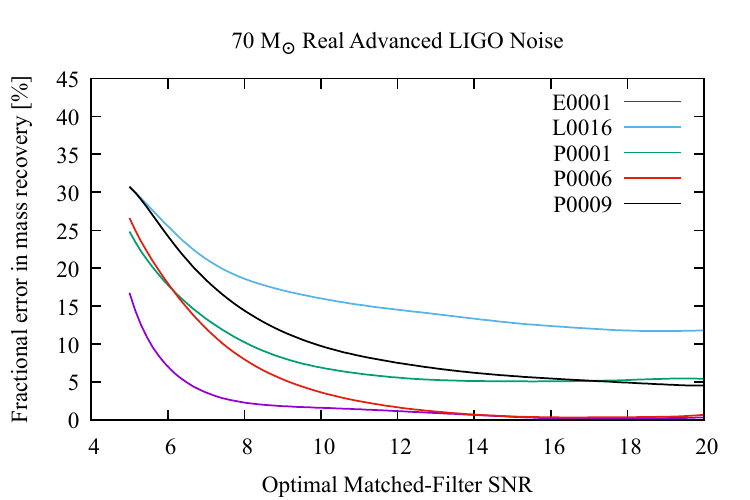}
  }
\centerline{
  \includegraphics[width=0.5\textwidth]{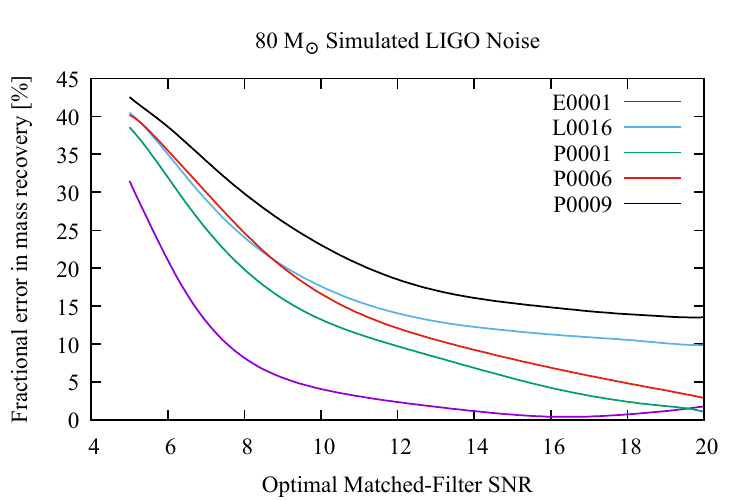}
 \includegraphics[width=0.5\textwidth]{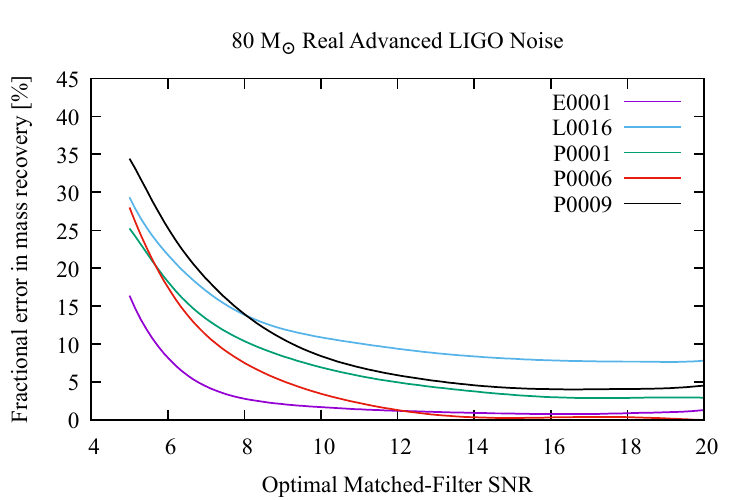}
  }
   \caption{The panels present mean percentage error of estimated total mass for each eccentric BBH population. These results were obtained using six hundred different noise realizations. Left panels: total mass recovery assuming simulated Gaussian noise for five different BBH populations, with three different total mass combinations. Right panels: as left panels but now using real advanced LIGO noise. We notice that in both cases, our neural network models can infer the total mass of the system better than 20\% for BBH systems with \(\textrm{SNR}\gtrsim15\). }
 \label{pe_total_mass}
 \end{figure*}

We have also utilized our neural network models to estimate the total mass of eccentric BBH mergers whose signals include higher-order waveform multipoles. We have considered several combinations of mass-ratio, total mass and eccentricity to explore the accuracy of our deep learning algorithms. In Figure~\ref{pe_total_mass} we present parameter estimation results for eccentric BBH populations with total mass \(M\in\{60\msun,\, 70\msun,\,80\msun\}\). These results represent mean percentage errors that we have obtained by averaging over six hundred different noise realizations. The left panels present results when using simulated LIGO noise, whereas the right panels present results using real advanced LIGO noise that is available at the \href{hhttps://www.gw-openscience.org/about/}{Gravitational Wave Open Science Center}~\cite{losc}.

\noindent Our results indicate that deep learning can constrain the total mass of a variety of eccentric BBH mergers with a fractional accuracy better than 20\% for GWs with \(\textrm{SNR}\gtrsim15\). We observe that systems with larger mass-ratio and eccentricities are more difficult to characterize. This is expected, since higher-order waveform multipoles introduce significant modifications to the ringdown evolution of these systems, and our deep learning algorithms have been trained with waveform datasets that describe non-spinning BHs on quasi-circular orbits. We also notice that our neural network model performs slightly better when using real advanced LIGO noise. In brief, while our deep learning algorithms are able to generalize to new types of signals, these results indicate that in order to improve detection and parameter estimation of eccentric BBH mergers, we need to design and train neural network models that are tailored for these type of systems, i.e., using training and testing data sets that describe eccentric BBH mergers. These investigations should be pursued in future work. 
\section{Conclusions}
\label{end}

We have quantified the importance of including higher-order waveform multipoles for GW searches of eccentric BBH mergers. Upon constraining the mass-ratio, eccentricity and binary inclination angles that maximize the contribution of higher-order waveform multipoles, we presented SNR calculations which indicate that \((\ell,\,\abs{m})\) modes become significant for BBHs with mass-ratios \(q\geq5\). Our findings indicate that for these types of eccentric BBH systems, \((\ell,\,\abs{m})\) modes play a more significant role for GW detection than for their quasi-circular counterparts. We then explored and quantified the significance of these results for the detection range within which ground-based LIGO-type detectors could observe these systems. Given the complex interplay between mass-ratio, eccentricity and the coupling of higher-order modes with spin-weight -2 spherical harmonics, we have prepared several visualizations to get insights into how eccentricity and mass-ratio modify these results~\cite{viz_homodes_dl}.

In addition to assessing the importance of \((\ell,\,\abs{m})\) modes for the detection of eccentric BBH mergers in terms of SNR calculations, it is also important to expand the existing signal-processing toolkit to search for and detect these complex GW signals. We have presented a preliminary analysis based on the use of neural network models to demonstrate that GWs that include higher-order waveform multipoles can be identified and extracted from simulated and real advanced LIGO noise, and that these algorithms can also constrain the total mass of eccentric BBH mergers. These findings are a first step towards the design, training and application of a new generation of deep learning algorithms that can probe much deeper signal manifolds. 

While this work has focussed on non-spinning BBHs that evolve on eccentric orbits, it is important to extend this study to astrophysically motivated scenarios in which BHs have non-negligible spin. A study of that nature will provide insights into the coupling of orbital eccentricity and spin. Previous studies have demonstrated that the signal manifolds describing non-spinning BHs on moderately eccentric orbits, and spinning BHs on quasi-circular orbits are degenerate~\cite{Huerta:2017a}. Thus, producing a NR waveform catalog of spinning BHs on eccentric orbits will be essential to better understand the physics of these GW sources. For instance, we have learned through analytical relativity and numerical relativity simulations that eccentricity shortens the length of waveform signals~\cite{Hinder:2010,ecc_catalog}. On the other hand, the dynamics of BBHs on quasi-circular orbits is significantly influenced by the spin configuration of their binary components. For instance, when the spin vectors are aligned with the total angular momentum, spin-orbit coupling effects delay the onset of merger; whereas 
in a spin anti-aligned configuration the merger phase is hastened. In different words, spin-orbit coupling also influences the length of waveform signals~\cite{Campanelli:2006uy}. These observations, combined with the results presented in this study, suggest that the inclusion of higher-order modes to describe asymmetric mass-ratio, spin-aligned BHs on moderately eccentric orbits will result in a measurable SNR increase. This is because, in this case, the net effect of spin is to extend the length of the waveform signals, which naturally translates into larger SNR values. We expect to see the opposite effect for spin anti-aligned and eccentric BBH mergers, since in this case both spin and eccentricity work together to shorten the waveforms, thereby leading to a reduction in SNR. These cases should be studied in further detail in the future to identify specific signatures that may enable us to tell apart the effects of eccentricity and spin in BBH mergers.

\section{Acknowledgements}

This research is part of the Blue Waters sustained-petascale computing project, which is supported by the National Science Foundation (awards OCI-0725070 and ACI-1238993) and the State of Illinois. Blue Waters is a joint effort of the University of Illinois at Urbana-Champaign and its National Center for Supercomputing Applications. We acknowledge support from NVIDIA, Wolfram Research, the NCSA and the SPIN Program at NCSA. NSF-1550514, NSF-1659702 and TG-PHY160053 grants are gratefully acknowledged. We thank the \href{http://gravity.ncsa.illinois.edu}{NCSA Gravity Group} for useful feedback. We thank one of the anonymous reviewers for providing constructive feedback regarding the importance of including higher-order modes to estimate the detectability range of eccentric BBH mergers.

\bibliography{references,dl_references}
\bibliographystyle{apsrev4-1}

\appendix
\section{Waveform Gallery}

\begin{figure*}[!htp]
\centerline{
 \includegraphics[width=0.5\textwidth]{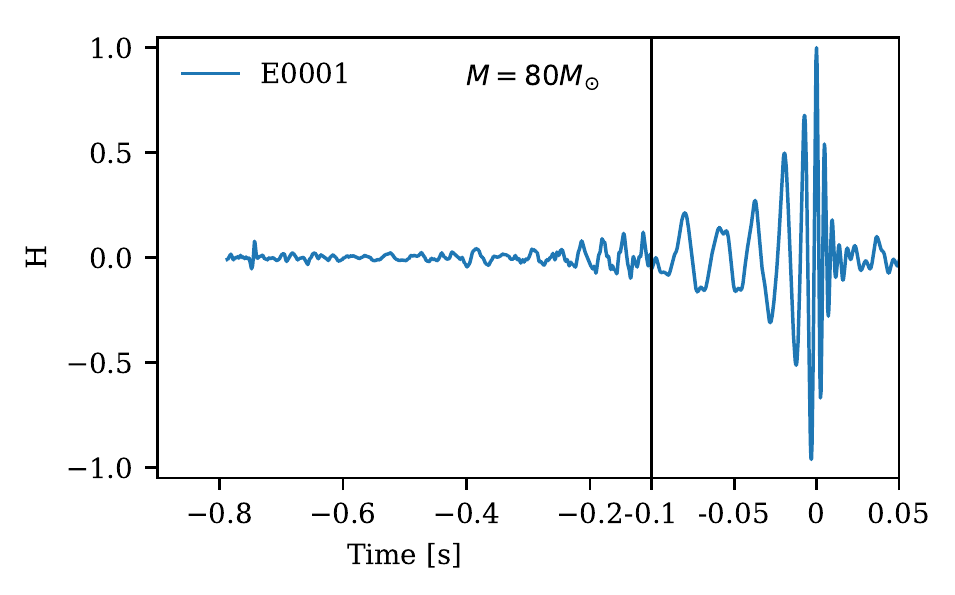}
  \includegraphics[width=0.5\textwidth]{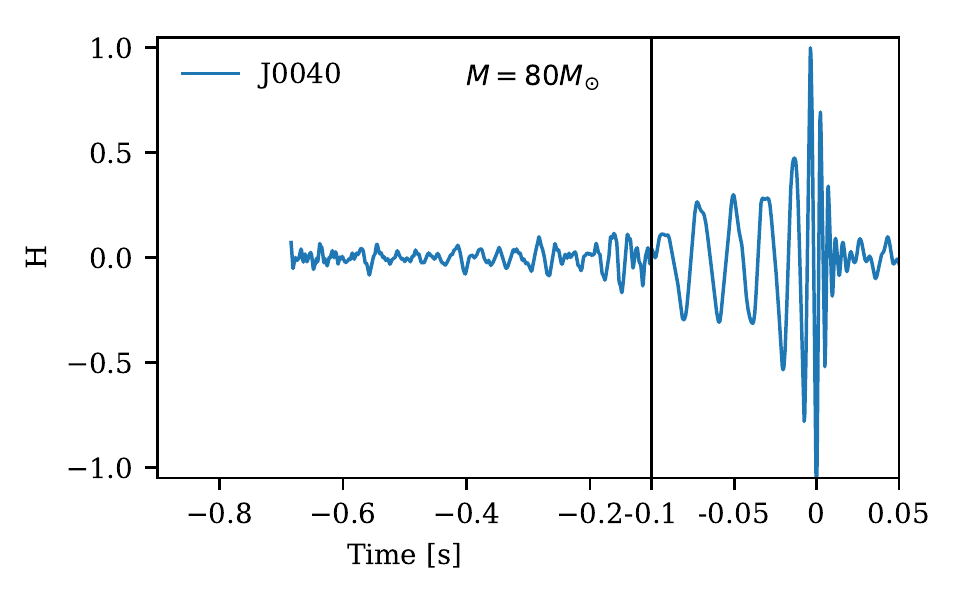}
  }
\centerline{  
 \includegraphics[width=0.5\textwidth]{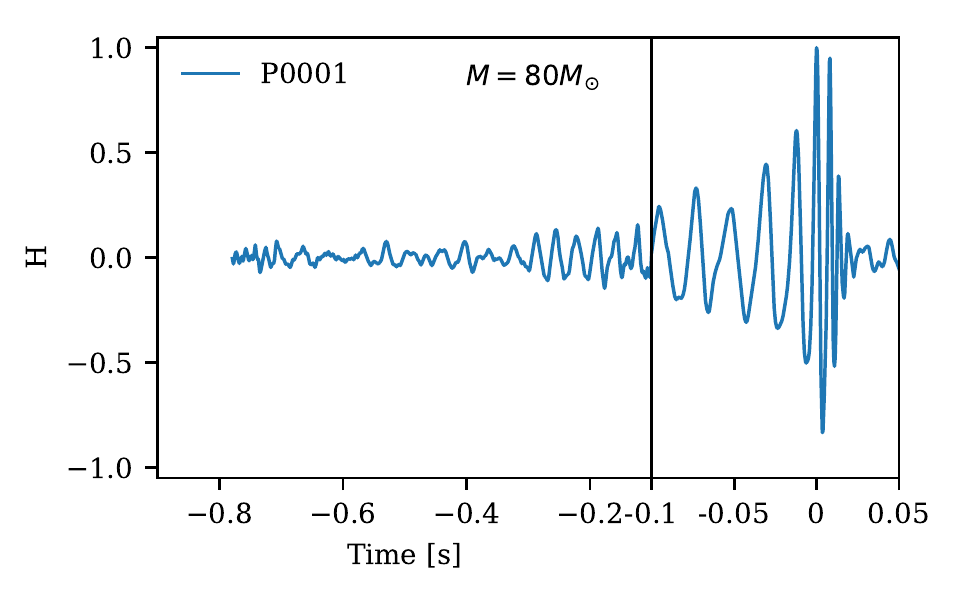}
  \includegraphics[width=0.5\textwidth]{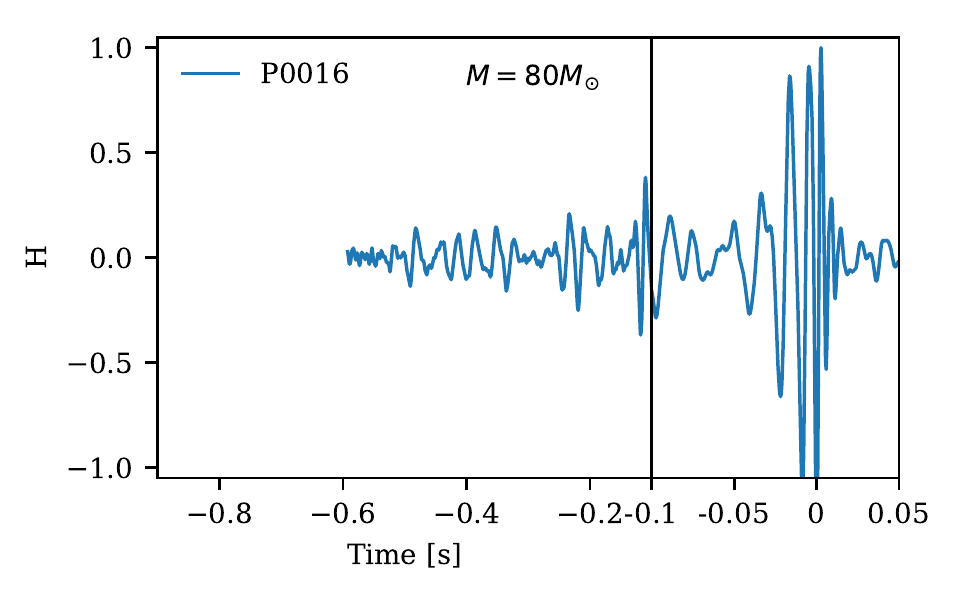}
  }
\centerline{
  \includegraphics[width=0.5\textwidth]{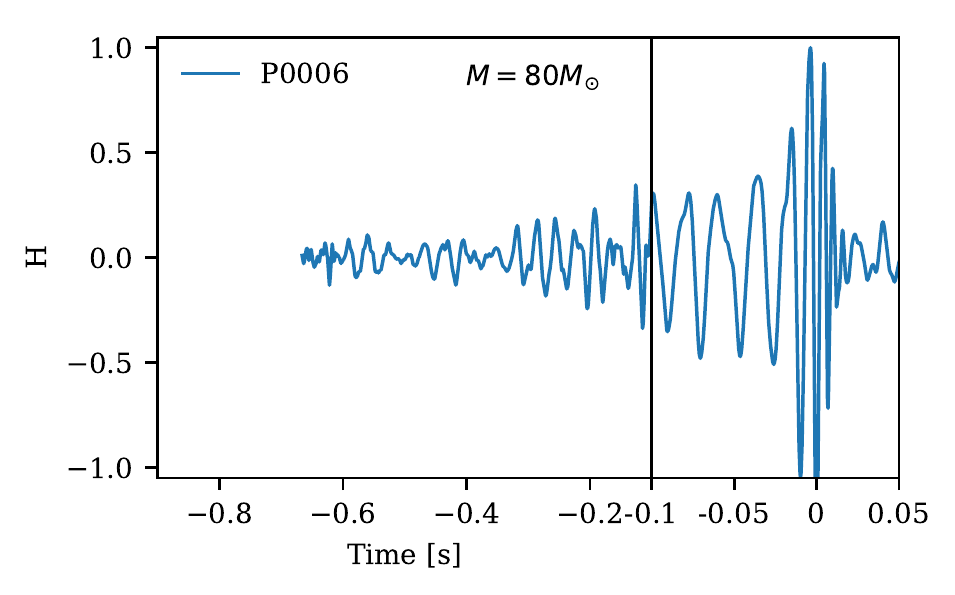}
 \includegraphics[width=0.5\textwidth]{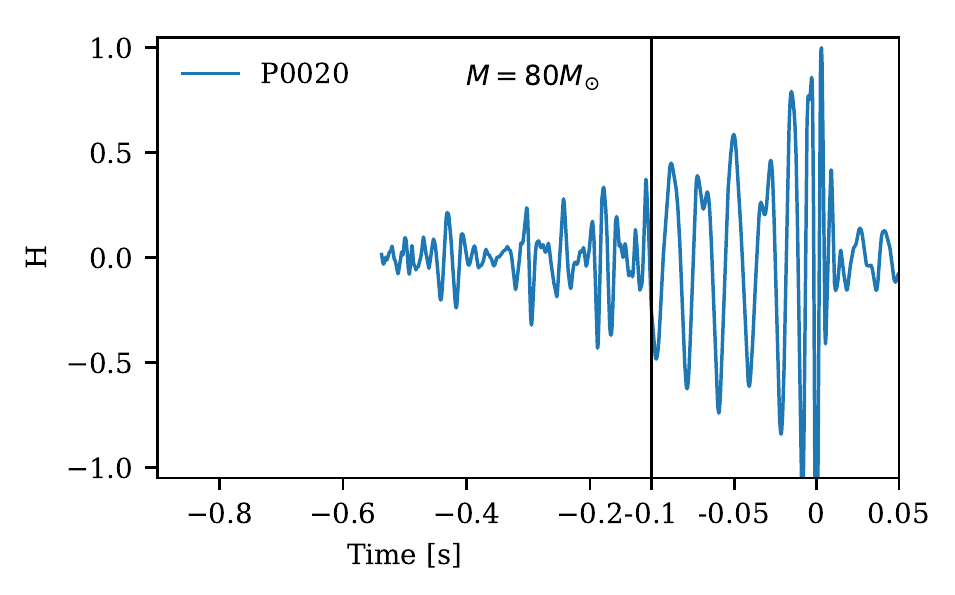}
  }
\centerline{
  \includegraphics[width=0.5\textwidth]{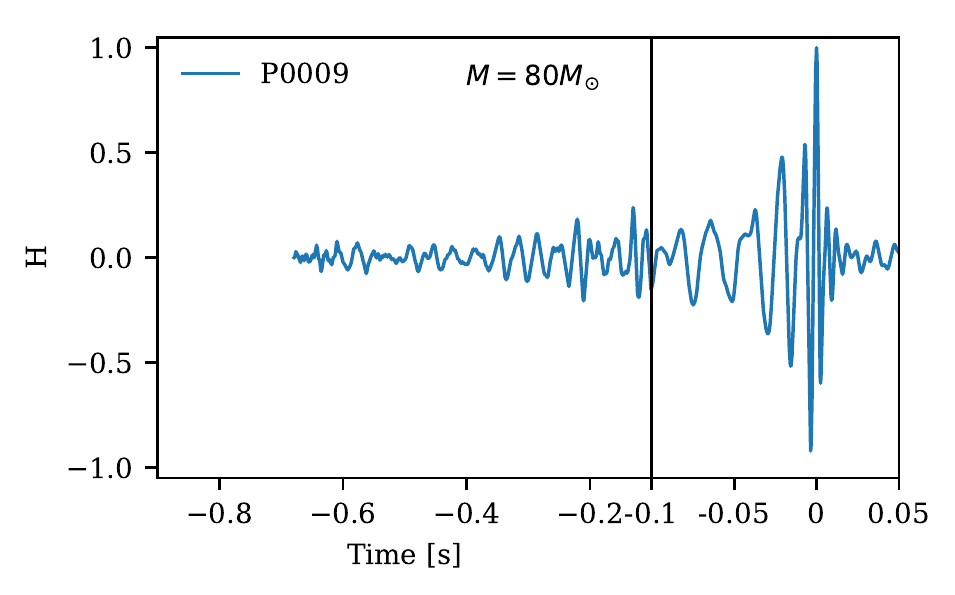}
 \includegraphics[width=0.5\textwidth]{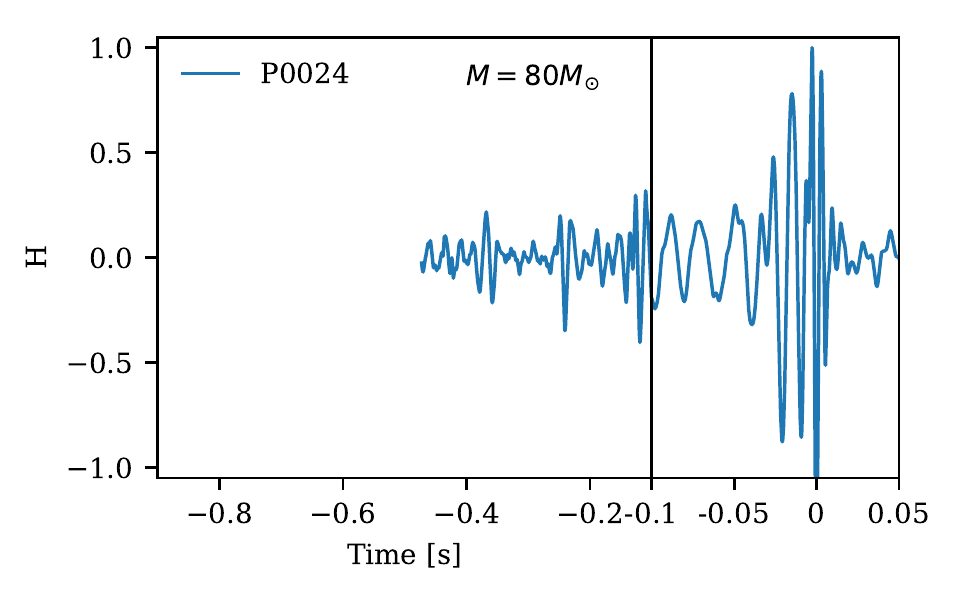}
  }
   \caption{Higher-order mode gravitational waves whitened with real advanced LIGO noise taken from its first observing run.}
 \label{waveforms}
 \end{figure*}

Figure~\ref{waveforms} presents a gallery of gravitational wave signals that we used in previous sections.

 \begin{figure*}[!htp]
\centerline{
 \includegraphics[width=0.5\textwidth]{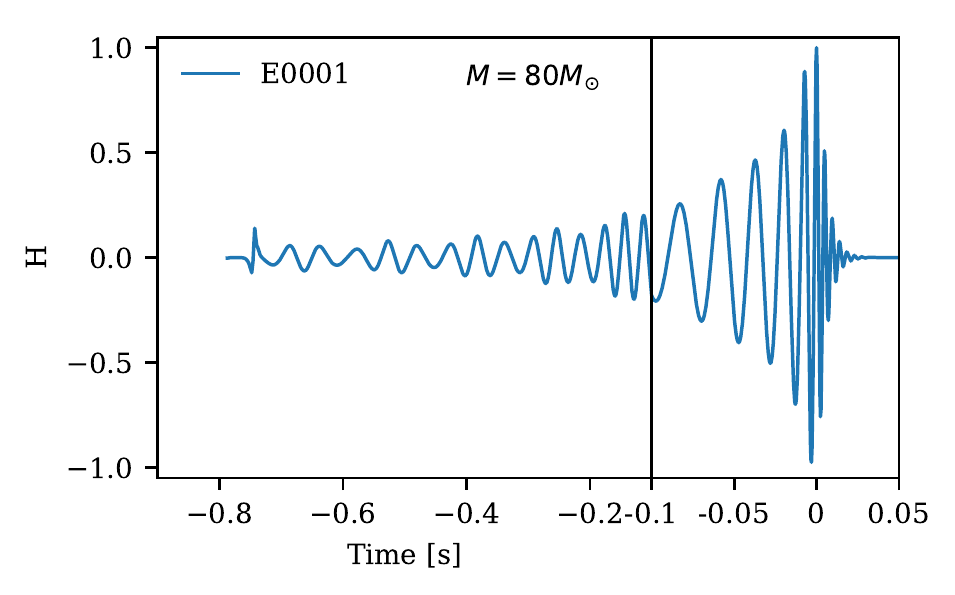}
  \includegraphics[width=0.5\textwidth]{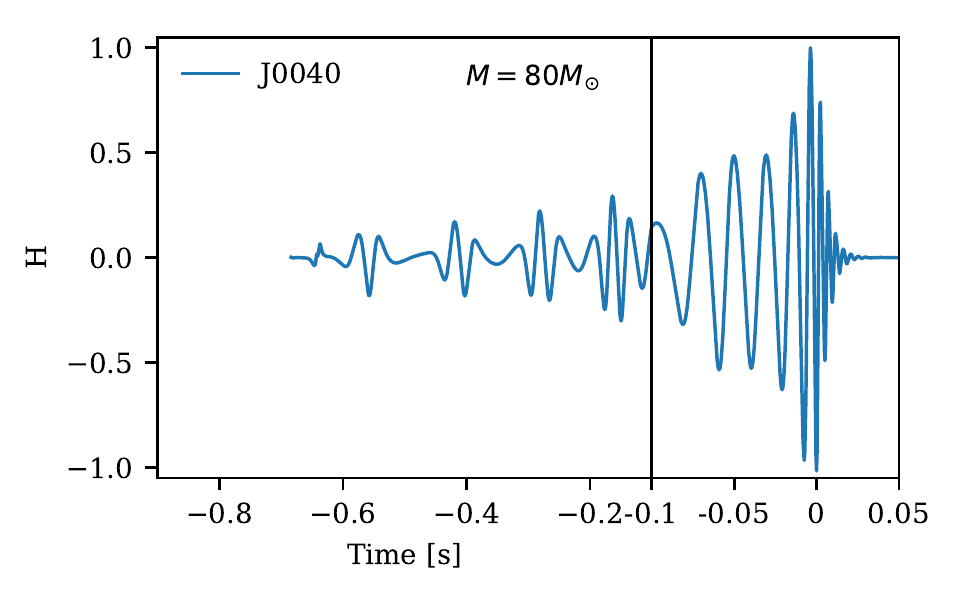}
  }
\centerline{  
 \includegraphics[width=0.5\textwidth]{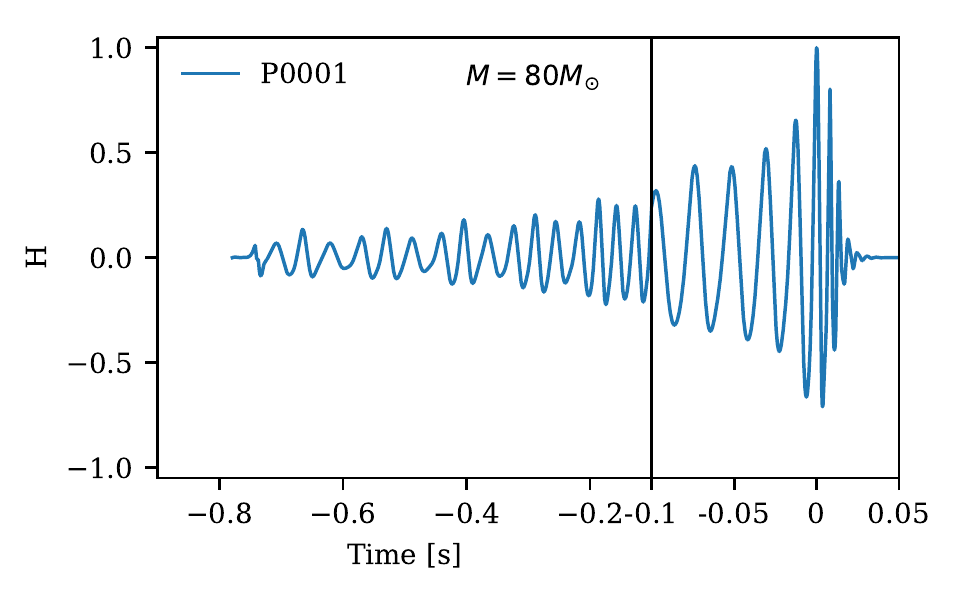}
  \includegraphics[width=0.5\textwidth]{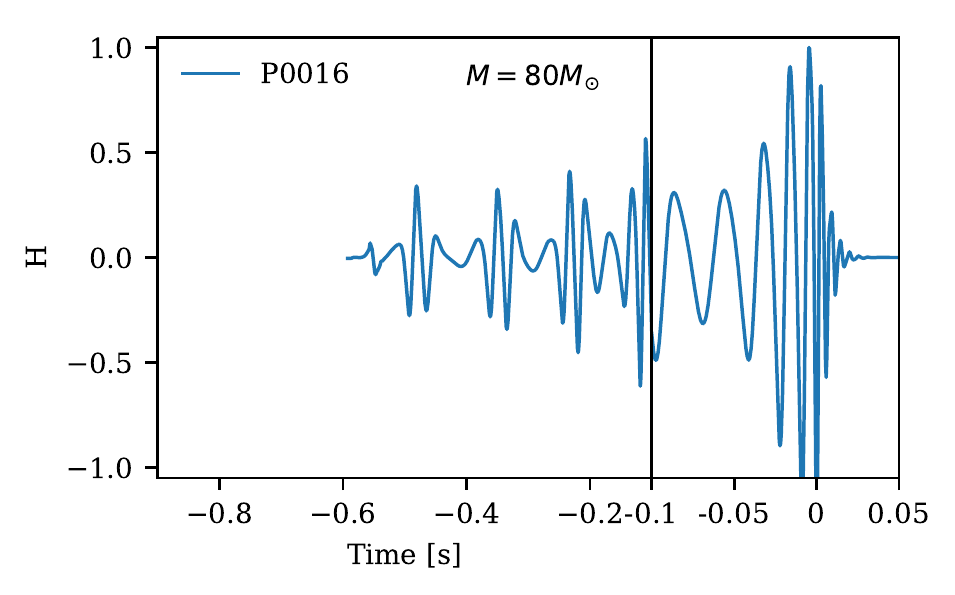}
  }
\centerline{
  \includegraphics[width=0.5\textwidth]{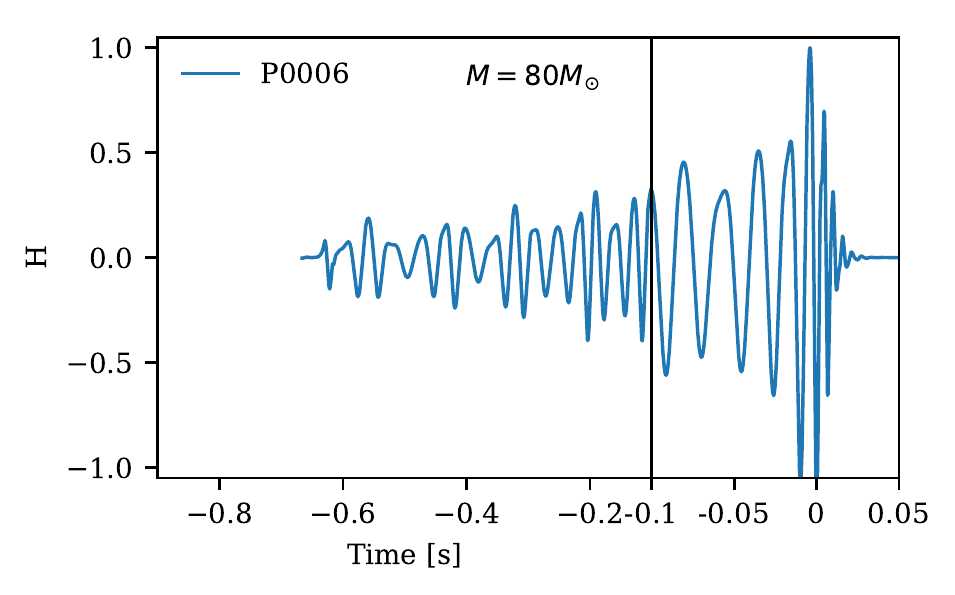}
 \includegraphics[width=0.5\textwidth]{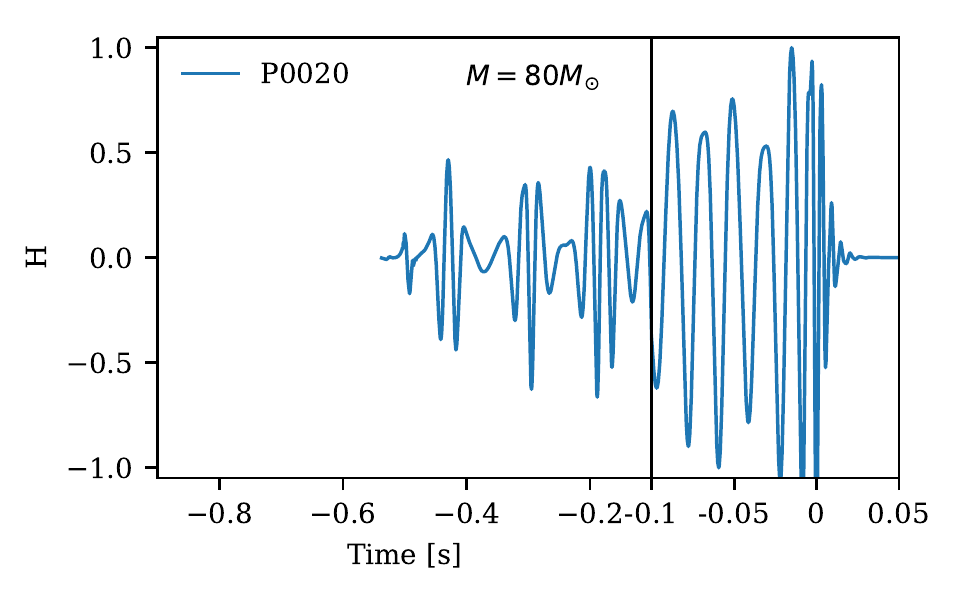}
  }
\centerline{
  \includegraphics[width=0.5\textwidth]{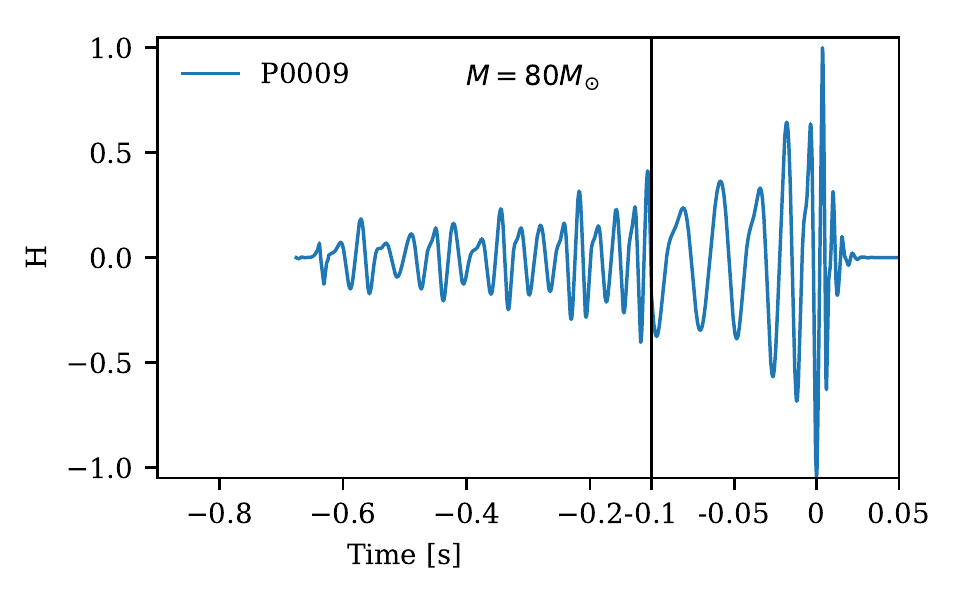}
 \includegraphics[width=0.5\textwidth]{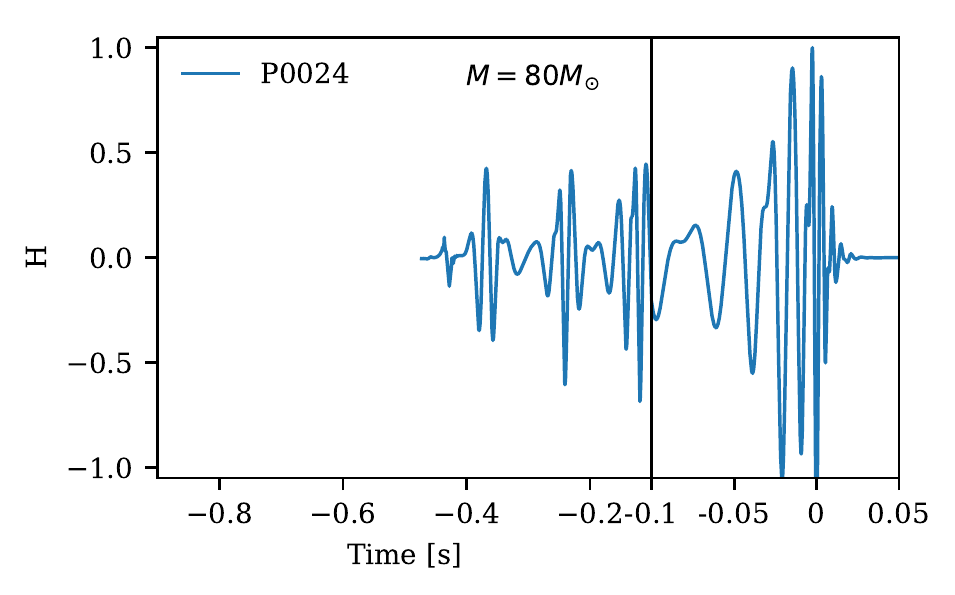}
  }
   \caption{Higher-order mode gravitational waves whitened with the target, simulated Zero Detuned High Power spectral density of the advanced LIGO detectors~\cite{ZDHP:2010}.}
 \label{waveforms_simulated}
 \end{figure*}
 
\end{document}